\documentclass[11pt,floatfix]{revtex4-1}
\usepackage{amsmath,amsfonts,amssymb}
\usepackage{graphicx}
\usepackage{float}
\usepackage[caption=false]{subfig}

\newcommand{\arctanh}[1]{\mathrm{arctanh}#1}

\newcommand{\Ham}{\mathcal{H}}
\newcommand{\B}{\beta}

\begin{document}
\title{Random Field Ising Model in two dimensions: Bethe approximation, Cluster Variational
Method and
message passing algorithms}

\author{Eduardo Dom\'{\i}nguez}
\affiliation{''Henri-Poincar\'e-Group'' of Complex Systems and
  Department of Theoretical Physics, Physics Faculty, University of
  Havana, La Habana, CP 10400, Cuba.}

\author{Alejandro Lage-Castellanos}
\affiliation{''Henri-Poincar\'e-Group'' of Complex Systems and
  Department of Theoretical Physics, Physics Faculty, University of
  Havana, La Habana, CP 10400, Cuba.}

\author{Roberto Mulet}
\affiliation{''Henri-Poincar\'e-Group'' of Complex Systems and
  Department of Theoretical Physics, Physics  Faculty, University of
  Havana, La Habana, CP 10400, Cuba.}

\date{\today}

\begin{abstract}
We study two free energy approximations (Bethe and plaquette-CVM) for the
Random Field Ising Model in two dimensions. We compare results obtained by these
two methods in single instances of the model on the square grid, showing the
difficulties arising in defining a robust critical line. We also
attempt average case calculations using a replica-symmetric ansatz, and
compare the results with single instances. Both, Bethe and plaquette-CVM
approximations present a similar panorama in the phase space, predicting long
range order at low temperatures and fields. We show
that plaquette-CVM is more precise, in the sense that predicts a lower critical
line (the truth being no line at all).  Furthermore, we give some insight on
the non-trivial structure of the fixed points
of different message passing algorithms.
\end{abstract}

\maketitle

\section{Introduction}
\label{sec:Intro}

Many of the thermodynamic properties of the Random Field Ising Model (RFIM) remained controversial for
more than 40 years. In particular the existence or the absence of a spin glass phase
in the model attracted a lot of attention\cite{DeDominicis,MezMona,Brezin,Almeida,Pastor,Fytas,Picco}.
Only recently we found confirmation that this spin glass phase is possible
only if the magnetization of the system is constrained to be fixed \cite{KrzRiccLen}. Without this constraint,
the RFIM possesses only a paramagnetic-ferromagnetic transition when $d>2$ or no transition at
all in $d\leq2$\cite{KrzRiccLen,KrzRiccSheLen,Imry_Ma_1,Imry_Ma_2}.

These results bring  new questions into the field. For example: Why standard perturbation
theory fails?  Could the spin-glass phase, deduced for the model with fixed
magnetization in the Bethe lattice \cite{KrzRiccLen}, have some role in the dynamical behavior of models
in finite dimensions? Do  message passing algorithms always converge in lattices
of finite dimensions as occurs in the ferromagnetic Ising model \cite{OurRCF}?

On the other hand, our current knowledge about the model makes it a good
starting point to improve the comprehension about the applicability and the capabilities
of techniques that have being used  to study this and other disordered systems.
Of special interest is  the analytical solution by M\'ezard and Parisi \cite{MP1,MP2}
of the  Viana-Bray model \cite{VB}) within a Replica Symmetry Breaking ansatz. Since then,
the field has progressed very fast. First, the ansatz was rapidly extended
to other models \cite{ScienceSAT,KS,Col}.
Then, it was soon recognized the connection between this approach and message  passing
algorithms, the well-known Belief Propagation (BP) algorithm \cite{Pearl} corresponds
to the Bethe approximation of the free energy of a specific model\cite{KABASAAD}.

To go from the Bethe approximation to methods that considered loops in the interaction
networks turned out to be a more difficult task \cite{MR,bolos1,bolos2,MooijKappen07,GKC10,zhou,zhou2,zhou3}.
An important step in that direction came from Yedidia and co-workers \cite{YFW05}
that described how to generalize the Cluster Variational Method (CVM) of Kikuchi
\cite{kikuchi}. The minimization of the CVM free energy can be achieved by the use
of a Generalized Belief Propagation (GBP) algorithm \cite{YFW05}, although the solution
found is always Replica Symmetric(RS).

More recently it was possible to merge the CVM with the Replica Symmetry Breaking
(RSB) ansatz\cite{tommaso_CVM}. Although, technically involved, the approach allowed
the study of the Edward-Anderson (EA) model in two dimensions.
In this system it
was possible to unveil the connection between the phase transition predicted
by the average case scenario and the properties of the Generalized Belief Propagation
algorithm in two dimensional lattices\cite{average_mulet}. Running standard BP for the Bethe
approximation in EA 2D one finds a paramagnetic solution at high temperature, as
expected. However, decreasing temperature BP finds, not one, but many fixed points
with non-zero local magnetizations.
Suggesting then, a transition from a paramagnetic to an spin glass phase that although
is not expected in thermodynamical grounds, is nevertheless predicted within the
approximation and connected with the performance of message passing algorithms
\cite{GBPGF,average_mulet}.
Furthermore, these low temperature fixed points are also correlated with
the metastable states of the Monte Carlo dynamics of the model \cite{our_metastables}.

In this work we present the equations describing the RFIM within the CVM
at the RS level. The goals are, on one side, to compare the predictions that can
be derived from the equations with our current knowledge about the model. This should
shed light on the limitations and advantages of the approximation itself. On the other,
to extend the connection already established within the EA model between these average
case equations and message passing algorithms. In particular, we will focus
on whether the CVM at the Bethe and plaquette level also predict a spin-glass phase,
and if this is connected with the behavior of message passing algorithms.

The rest of the work is organized as follows.  In the next section \ref{sec:Mod}, we
present the model. In section \ref{sec:GBP} we obtain message passing equations for
the single instances of the model, both at the Bethe  (BP) and the plaquette level of CVM (GBP).
The phase diagram obtained by implementing these belief propagation algorithms is discussed in \ref{sec:Single_Inst}.
In section \ref{sec:Ave} we derive average case predictions at replica symmetric
level for both approximations. We summarize and discuss our findings and conclusions in \ref{sec:Conc}.

\section{The Model}
\label{sec:Mod}
 The Random Field Ising Model is a natural extension of the standard Ising
ferromagnet to consider the disorder
of real solids. Due to a number of reasons, spins in a crystalline lattice
are under the influence of small local magnetic fields;
intensity and direction of these fields varies rapidly from one site to
another uncorrelatedly. This situation can be modeled
in a simplified way by considering the following Hamiltonian:
\[
\Ham (\sigma) = -  J  \sum_{\langle i,j\rangle}\sigma_i \sigma_j - \sum_i
h_i \sigma_i
\]
where the first sum runs over all couples $\langle i,j\rangle$  of neighboring  spins
(first neighbors on the lattice) and the second over all sites (spins). We will deal in this
work with $N=L\times L$ spins in a bidimensional square lattice, but extensions to more dimensions is conceptually
straightforward, although probably numerically cumbersome. The magnetic exchange
constants between spins is fixed to $J>0$ (more precisely $J=1$)  and the spins  $\sigma_i=\pm 1$ are the dynamic variables.
The disordered local field $h_i$ are a set of random variables
generally drawn from a zero-mean bimodal or Gaussian distribution.
In our case, a bimodal distribution will be used:
\begin{equation*}
P_h\left[h_i \right]= \dfrac{1}{2} \left( \delta(h_i-H) + \delta(h_i+H)
\right)
\end{equation*}

Local fields introduce what is called quenched disorder \cite{MPV}: fields and spins
are considered to vary in completely different time scales in real solids. Thus, an instance of this
model corresponds to a particular realization of the fields. We will use the field intensity
$H$ as one of the model parameters.

The statistical mechanics of the RFIM model, at a temperature $T=1/\B $,
is given by the Gibbs-Boltzmann distribution
\[P(\sigma) = \frac{e^{-\B\Ham(\sigma)}}{Z} \quad \mbox{where }\quad Z=\sum_{\sigma} e^{-\B \Ham(\sigma)}
\]
is a normalization constant, called partition function since it depends on the temperature $T=1/\B$
and all other relevant parameters of the model, like $H$ for instance.

Finding the exact free energy of a particular sample of our model is quite
difficult, because the partition function involves an exponential number of computations,
 being only accessible to very small systems. However, the interesting limit
is the thermodynamic one ($N\to \infty$), and from a physics perspective we are
mostly interested in a description of the ensemble of instances, not in a particular
one. Using the replica method for models in fully connected or very sparse random topologies
 (see \cite{MPV} for a primer), the structure of equilibrium states for the average
case can be described in a well defined hierarchy of approximations. The whole
approach is based on the idea that the free energy is a self-averaging quantity, so
averaging over disorder must match results for real samples when $N\to \infty$.
An exact solution to disordered models in finite dimensions, however, remains immune
to analytic results, and usually is treated by simulations or approximations, like the ones
described next.

\section{GBP equations for plaquette-CVM}
\label{sec:GBP}

The fundamental idea behind Generalized Belief Propagation \cite{YFW05}, or Cluster
Variational Method \cite{yedidia,pelizzola05}, or Kikuchi approximation \cite{kikuchi}
to the free energy of a model
is to replace the exact functional free energy $F[P(\sigma_1,\sigma_2,\ldots,\sigma_N)]$
by an approximation consisting on the sum of local (small regions) free energies:
\[ F\simeq F_{\mbox{\scriptsize CVM}} = \sum_{R\in\mathcal{R}} c_R F_R[P_R(\sigma_R)].
\]
Since the set of regions is in principle arbitrary, some regions might contain
smaller regions, and therefore an appropriately chosen counting number $c_R$ is
used to avoid overcounting free energy contributions \cite{yedidia}. Only in
very particular cases a partition like this remains exact (like in the Bethe
approximation on tree like topologies) or even an upper bound of the real free
energy \cite{yedidia}.

Ideally the distributions minimizing $F_{\mbox{\scriptsize CVM}}$ should coincide
with the marginals of the exact Boltzmann distributions
$P_R(\sigma_R) = \sum_{\sigma \setminus \sigma_R} P(\sigma_1,\sigma_2,\ldots,\sigma_N)$, but generally (and hopefully)
they will only be an approximate of them, and therefore are usually called {\it beliefs}
and represented by $b_R(\sigma_R)$, leaving $P_R(\sigma_R)$ for the exact marginals.
\begin{figure}[ht!]
 \centering
 \includegraphics[width=0.7\textwidth,keepaspectratio=true]{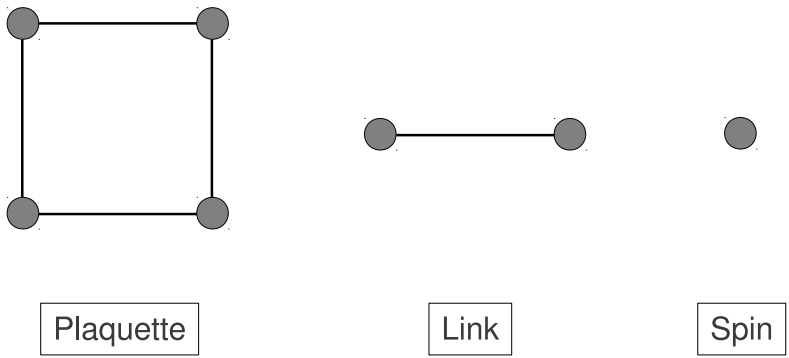}
 \caption{Regions for CVM approximation} \label{fig:regiones}
\end{figure}

Though very general, there are specific ways to define the set of regions $\mathcal{R}$
for approximating the free energy. According to the CVM prescription \cite{kikuchi,yedidia},
for instance, this set is formed from a primary set of larger regions
$\mathcal{R}_0$. Then adding to them the set $\mathcal{R}_1$, formed by the intersections of every
pair of regions in $\mathcal{R}_0$. Then adding the intersections of regions in
$\mathcal{R}_1$ and so on and so forth until no intersections are left, obtaining
$\mathcal{R}=\mathcal{R}_0 \cup \mathcal{R}_1 \cup \mathcal{R}_2 \ldots$.

Bethe approximation is just a particular case of CVM where the larger regions considered
are the interacting pairs of spins (links). In this work we use Bethe approximation,
and, to go beyond, we also use the approximation starting from plaquette regions.
A plaquette region is formed by the four spins and links lying on a basic square cell of the
lattice (see Fig. \ref{fig:regiones}), and has the advantage over Bethe approximation
of including the shortest loops in the graph. It is known that loops are the cause of
failure of Bethe approximation. Two adjacent plaquettes overlap in a link region, formed by
two neighbor  spins and the interaction between them. Two link regions with a common spin
overlap in a spin region. Therefore, our approximation contains contributions of free energies
from all plaquettes ($P$), links ($L$) and sites ($i$) (Fig. \ref{fig:regiones}). The free energy approximated
with this set of regions is usually  called Kikuchi approximation \cite{kikuchi,pelizzola05}:
\[F_{\mbox{\tiny CVM}} [\{b_P\},\{b_L\},\{b_i\}]= \sum_{P} F_P[b_P(\sigma_a,\sigma_b,\sigma_c,\sigma_d)]
-\sum_{L} F_L[b_L(\sigma_a,\sigma_b)] +\sum_{i} F_i[b_i(\sigma_i)] \]

\begin{figure}
\subfloat[Link to spin]{\label{fig:message_update_equations_link}
\includegraphics[keepaspectratio=true,width=0.49\textwidth]{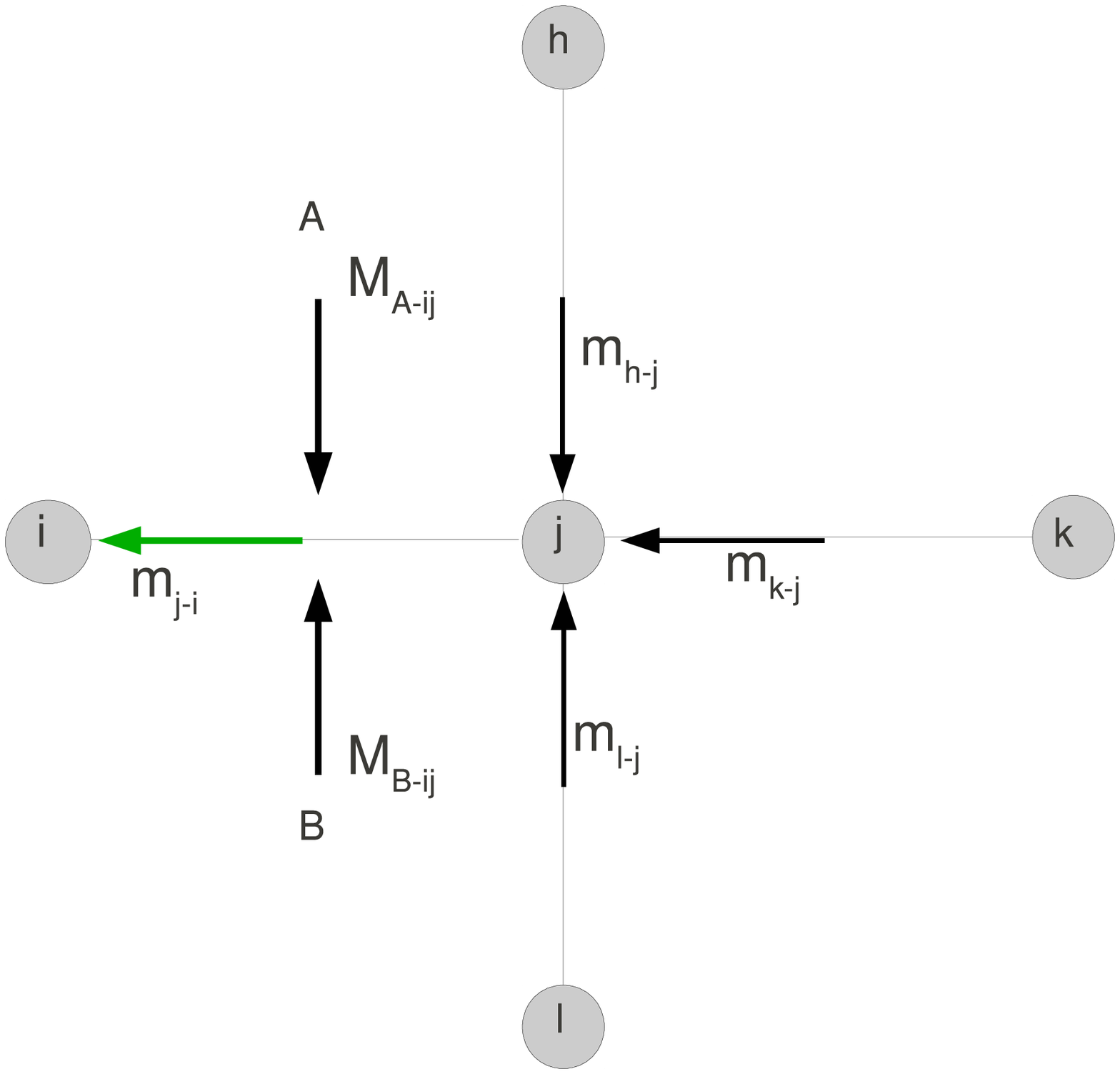}}
\subfloat[Plaquette to link]{\label{fig:message_update_equations_plaq}
\includegraphics[keepaspectratio=true,width=0.495\textwidth]{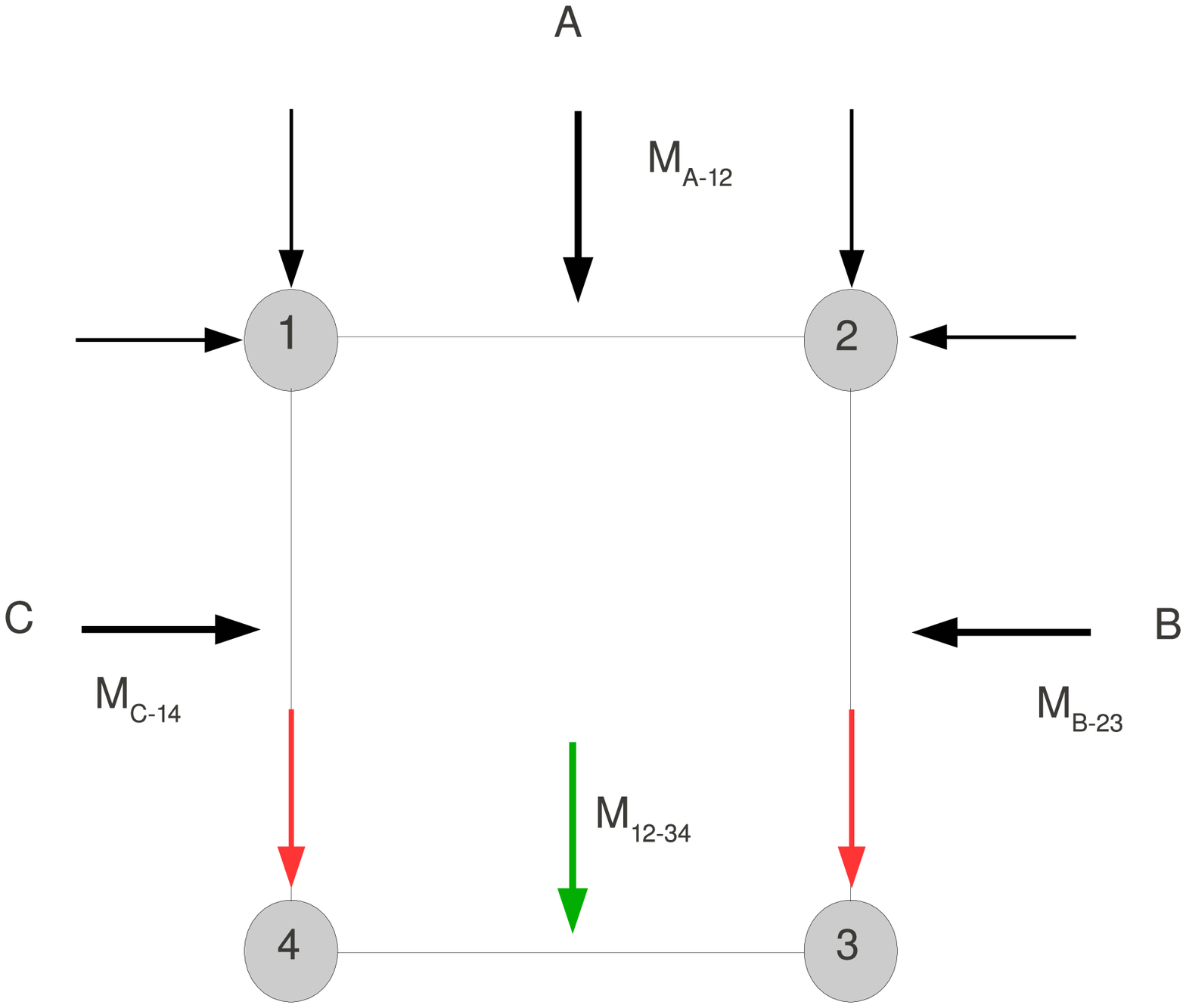}}
\caption{Schematic representation of message update equations. Green messages are
updated using black and red ones. Red messages appear on the LHS of the respective equation.}
\label{fig:message_update_equations}
\end{figure}

Minimization of this free energy with respect to the beliefs $\{b_P\},\{b_L\},\{b_i\}$
has to be done while keeping the consistency among them, in the sense that distributions on larger regions
have to marginalize onto the smaller regions that are part of them. This constrained
minimization is implemented via Lagrange multipliers $M_{P\to L}$ and $m_{L\to i}$ enforcing
plaquette to link and link to spin consistency respectively. The self consistent
equations for the Lagrange multipliers are interpreted as message passing equations.
The message from link $(i,j)$ to spin, say, $i$ is updated according to:
\begin{equation}
 \label{eqn:link2spinUpdate}
 m_{(i,j)\rightarrow i}(\sigma_i)\propto \sum_{\sigma_j} \psi_{ij}(\sigma_i,\sigma_j) \prod_{P\in \mathcal{P}[(i,j)]}  M_{P\rightarrow (i,j)}(\sigma_i,\sigma_j) \prod_{L\in \mathcal{L}(j)\setminus (i,j)} m_{L\rightarrow j}(\sigma_j)
\end{equation}
where $\mathcal{P}[(i,j)]$ is the set of regions of whom $(i,j)$ is a
child. In  Fig. \ref{fig:message_update_equations_link} it is shown the configuration
of all messages involved in updating $m_{(i,j)\rightarrow i}(\sigma_i)$.
The interactions appear as $\psi_{ij}(\sigma_i,\sigma_j)=\exp(-\B E_{ij}(\sigma_i,\sigma_j))$, where $E_{ij}(\sigma_i,\sigma_j)=-J\sigma_i\sigma_j-h_j\sigma_j$.
The product over link-to-spin messages runs on $\mathcal{L}(j)$, which
is the set of links having the site $j$ as a child, excluding $(i,j)$. In the case
shown, $\mathcal{L}(j)\setminus (i,j)$ includes $(h,j)$, $(k,j)$ and $(l,j)$.

For messages from the square plaquette $A=(ijkl)$ to link $(i,j)$ (see Fig. \ref{fig:message_update_equations_plaq})
we have the update equation (\ref{eqn:updateplaqtolink}) that is similar in structure.
This time, however, on the left hand side appears the product of two link messages
that are internal to the plaquette. Normally it is advisable (for convergence issues)
to update these first before updating $M_{A\to(i,j)}$.
\begin{eqnarray}
\nonumber
\lefteqn{
 M_{A \rightarrow (i,j)}(\sigma_i,\sigma_j)m_{(i,l)\rightarrow i}(\sigma_i)
m_{(j,k)\rightarrow j}(\sigma_j)\propto
\left\lbrace \sum_{\sigma_k, \sigma_l} \psi_{ijkl}(\sigma_i,\sigma_j,\sigma_k,\sigma_l) \right.}\\
&&
\label{eqn:updateplaqtolink}
\left.
\times \left(
\prod_{L\in \mathcal{L}_P[A]\setminus (i,j)}
\prod_{P\in \mathcal{P}[L]\setminus A} M_{P\rightarrow L}(\sigma_L) \right)
\left(
\prod_{n \in \{k,l\}} \prod_{L\in \mathcal{L}_S[n] \wedge L\notin \mathcal{L}_P[A]} m_{L\rightarrow n}(\sigma_n) \right)
\right\rbrace \quad
\end{eqnarray}
Here, $\mathcal{L}_P[A]$ is the set of links in plaquette $A$, $\mathcal{L}_S[n]$ represents the links that are parent to spin $n$
and $\mathcal{P}[L]$ is the collection of parent plaquettes to $L$.

Messages are the usual representation of the Lagrange multipliers in the context
of belief propagation algorithms. However, physically it is nicer to think in terms
of {\it cavity fields} \cite{MP1,MP2}, acting from one part of the system onto another. In terms of
these fields, messages are recovered as
\begin{equation}
m_{(i,j)\to i}(\sigma_i)  = \exp{\beta u_{Li} \sigma_i} \qquad
M_{A \rightarrow (i,j)}(\sigma_i,\sigma_j) = \exp{\beta (U \sigma_i \sigma_j +u_i \sigma_i + u_j \sigma_j)}
\label{eqn:m_to_u}
\end{equation}
Only one parameter $u_{Li}$
is enough to represent $m_{L\to i}$ messages, while for  $M_{P \to (ij)}$ we need three:
$U_{Pij}$, $u_{Pi}$ and $u_{Pj}$ \cite{tommaso_CVM,dual,GBPGF}. The update equations for the cavity fields are obtained
with little effort. Below we write them using the labeling of Fig. \ref{fig:message_update_equations}.
\begin{eqnarray}
 \hat{u}_{j \to i}(\#)&=&\dfrac{1}{\B}\arctanh{[\tanh{\B \hat{J}}\tanh{\B \hat{h}_j}]}
+ u_{P_A\to i}  + u_{P_B\to i}  \label{eqn:hat_u} \\
 &\mbox{where}& \nonumber\\
 \nonumber
 \hat{h}_j&=&h_j + u_{l\to j} + u_{k \to j} + u_{h\to j} + u_{P_A \to j} + u_{P_B\to j} \nonumber \\
 \hat{J}&=&J+ U_{A\to ij}+ U_{B\to ij} \nonumber
\end{eqnarray}
Fields define a new effective $\hat{h}_j$ and coupling constant $\hat{J}$.
For plaquette-to-link
fields we get
\begin{eqnarray}
 \hat{U}_{A\to ij}(\#)&=&\dfrac{1}{4\B} \ln \dfrac{K(1,1)K(-1,-1)}{K(-1,
1)K( 1,-1)} \label{eqn:hat_Uuu} \\
 \hat{u}_{A\to i}(\#)&=&  u_{D\to i} -u_{l\to i} + \dfrac{1}{4\B} \ln \dfrac{K(1,1)K(1,-1)}{K(-1, 1)K(-1,-1)} \nonumber \\
 \hat{u}_{A\to j}(\#)&=&  u_{B\to j} -u_{k\to j} + \dfrac{1}{4\B} \ln \dfrac{K(1,1)K(-1, 1)}{K( 1,-1)K(-1,-1)} \nonumber
\end{eqnarray}
where
\begin{equation*}
 K(\sigma_i,\sigma_j)=\sum_{(\sigma_k,\sigma_l)} \exp \B \left( \hat{U}_{D\to
il}\sigma_i\sigma_l + \hat{U}_{C\to kl}\sigma_k\sigma_l + \hat{U}_{B\to jk}\sigma_j\sigma_k + \hat{h}_k \sigma_k +\hat{h}_l \sigma_l \right)
\end{equation*}

The set of self consistent equations defining messages can be solved by a fixed point iteration. This procedure
is known as a \textit{message passing algorithm}. Standard Belief Propagation (BP) is just a special case for the Bethe
approximation. Many details about the implementation of BP and GBP can be found in  \cite{GBPGF}. If the algorithm
happens to find a fixed point (convergence is not guaranteed), you can recover the local approximated marginals $b_R(\sigma_R)$ in terms
of the messages, just like with any standard Lagrange multiplier minimization.

Now that we have presented in broad terms the essential features of BP and GBP, we can move to discussing the results when
running these fixed point methods on singular instances of 2D RFIM.

\section{Results for single instances}
\label{sec:Single_Inst}

Region graph approximation to the free energy and the corresponding message passing algorithms
give numerical (estimate) values to all thermodynamic quantities like free energy, energy or entropy.
Usually these approximations are compared with Monte Carlo simulation results, or exact predictions (when
available). We are most interested in the phase diagram of the model.

Two relevant parameters characterize the random field Ising model: the temperature $T=1/\beta$ and the
external field intensity $H$. Local fields $h_i = H \tilde h_i$ are scaled by $H$, and
$\tilde h_i \in \{-1,1\}$ with equal probability. The model with $H=0$ is no other than
the usual Ising ferromagnet.

\subsubsection*{Phase Diagram}
We know in advance the dominant thermodynamical phase of some particular points in the $(T,H)$ plane.
For instance, at any finite field intensity $H$, and high enough temperature $T$, the free energy
minimization is dominated by the entropic term, and therefore spins are mostly uncorrelated
(true when $T\to\infty$) with vanishing local  magnetizations, $m_i\simeq \tanh \beta h_i \sim \B H =H/T$,
and with global zero magnetization $M = \frac 1 N \sum_i m_i \sim N^{-1/2}H/T$ given the random origin
of the fields $\tilde h_i$.

At zero temperature, the free energy is dominated by the minimization of the energetic term,
and at least for $H>4$, the interaction with the 4 surrounding spins at any given site cannot overcome
the interaction with the local external field. In such case, the ground state of the system
is trivially $\forall_i \sigma_i = \mbox{Sign}(\tilde h_i)$. The system is frozen and again global
magnetization is zero (except for $1/\sqrt{N}$ fluctuations). Let's call this phase
paramagnetic in spite of spins been magnetized.
In this sense, paramagnetic refers to the fact that no long range order exists, and the frozenness
of the spin is not a collective phenomena but a result of their interaction with external local fields.

Therefore, in both extremes, high $T$ and high $H$, the system is paramagnetic, and this is correctly
reproduced by both, Bethe and CVM approximations, on single instances (see below). To signal the onset of long range correlations we have used the global magnetization as the order parameter and a threshold value
$|M| > M_{th}$.
The value of this threshold can be fixed according to the following reasoning. For a given temperature,
long range order, if present, is destroyed for large $H$. In this regime, magnetization is of order
$m\simeq \frac{1}{\sqrt{N}}=\frac{1}{L}$. If the system remains paramagnetic when decreasing $H$, magnetization should decrease too.
 On the contrary, if for lower $H$ states with $m \gg \frac{1}{L}$ are found,
 it can be interpreted as the emergence of order in the system. In practice, a value of $M_{th}=\frac{5}{L}$ fits known
 results for $H=0$, namely the para-ferro transition point of the Ising model.

\begin{figure}[!htb]
 \includegraphics[keepaspectratio=true,width=0.75\textwidth]{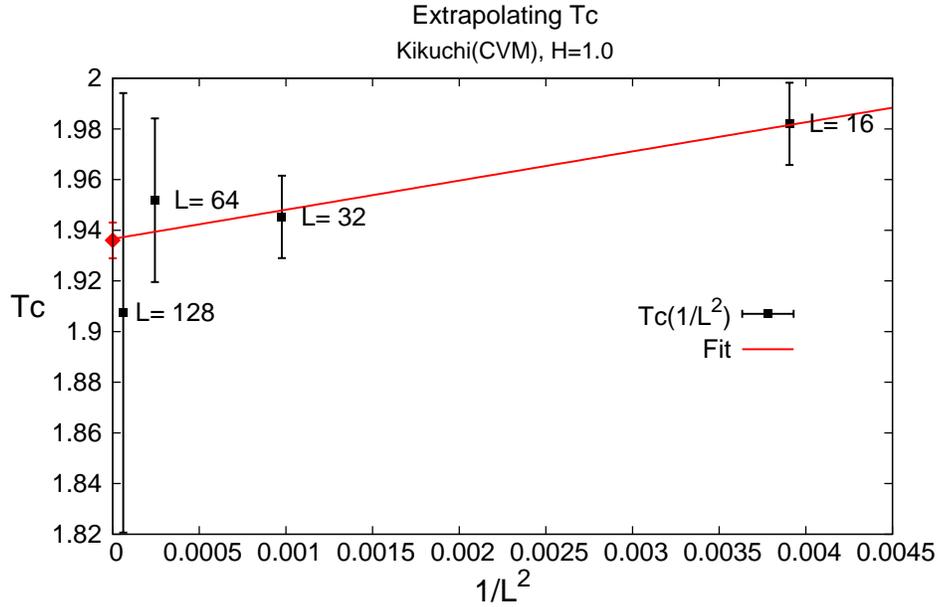}
 \caption{For a fixed field intensity, different $T_c$ values are found depending on the size of
the lattice. Here we assume $|T_c(L)-T_c(\infty)|\propto \dfrac{1}{L^2}$ and make a linear extrapolation for $L\to \infty$.
This procedure is repeated for each $H$ value in the CVM and Bethe approximation.}
 \label{fig:kikuchi_tc_extrapol}
\end{figure}

To unveil the phase diagram in the thermodynamic limit from simulations on finite size instances we considered that, for a given value of the $H$ field, the mean value of the critical temperature over many samples depends
on the system size, $T_c=T_c(L)$. In order to obtain an estimate of the value in the $L\to\infty$ limit,
we propose $|T_c(L)-T_c(\infty)|\propto \dfrac{1}{L^2}=\dfrac{1}{N}$ and extrapolate (see figure
\ref{fig:kikuchi_tc_extrapol}).

\begin{figure}
\subfloat[ Critical lines for L=16,32,64.]{\label{fig:bethe_si_all}
 \includegraphics[keepaspectratio=true,width=0.49\textwidth]{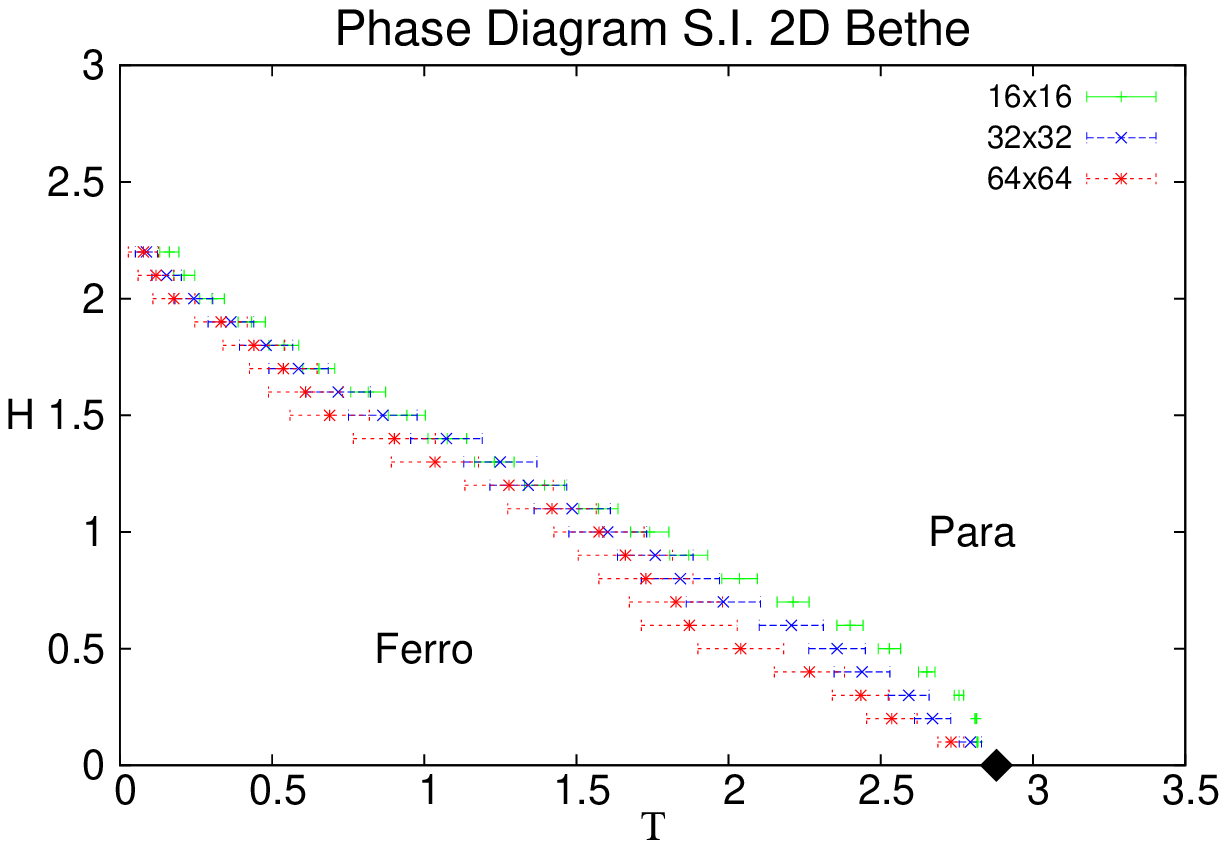}}
\subfloat[Extrapolated $T_c(L)$ for $L \rightarrow \infty$.]{\label{fig:bethe_si_extrapolated}
 \includegraphics[keepaspectratio=true,width=0.49\textwidth]{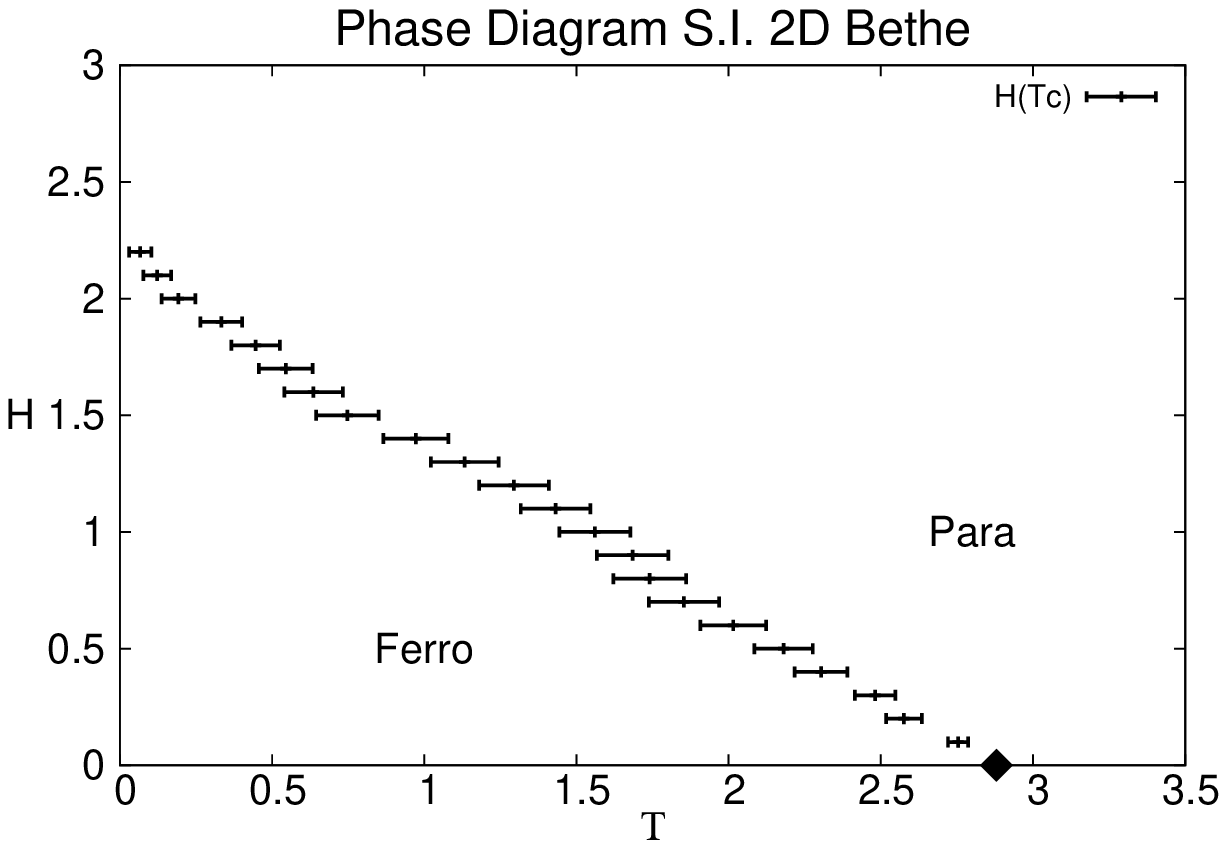}}
\caption{Bethe(BP) approximation. Note in both plots that for $H\rightarrow 0$ the result for the Ising magnet is recovered.}
\label{fig:bethe_si}
\end{figure}

\begin{figure}
\subfloat[ Critical lines for L=16,32,64,128.]{\label{fig:kikuchi_si_all}
\includegraphics[keepaspectratio=true,width=0.49\textwidth]{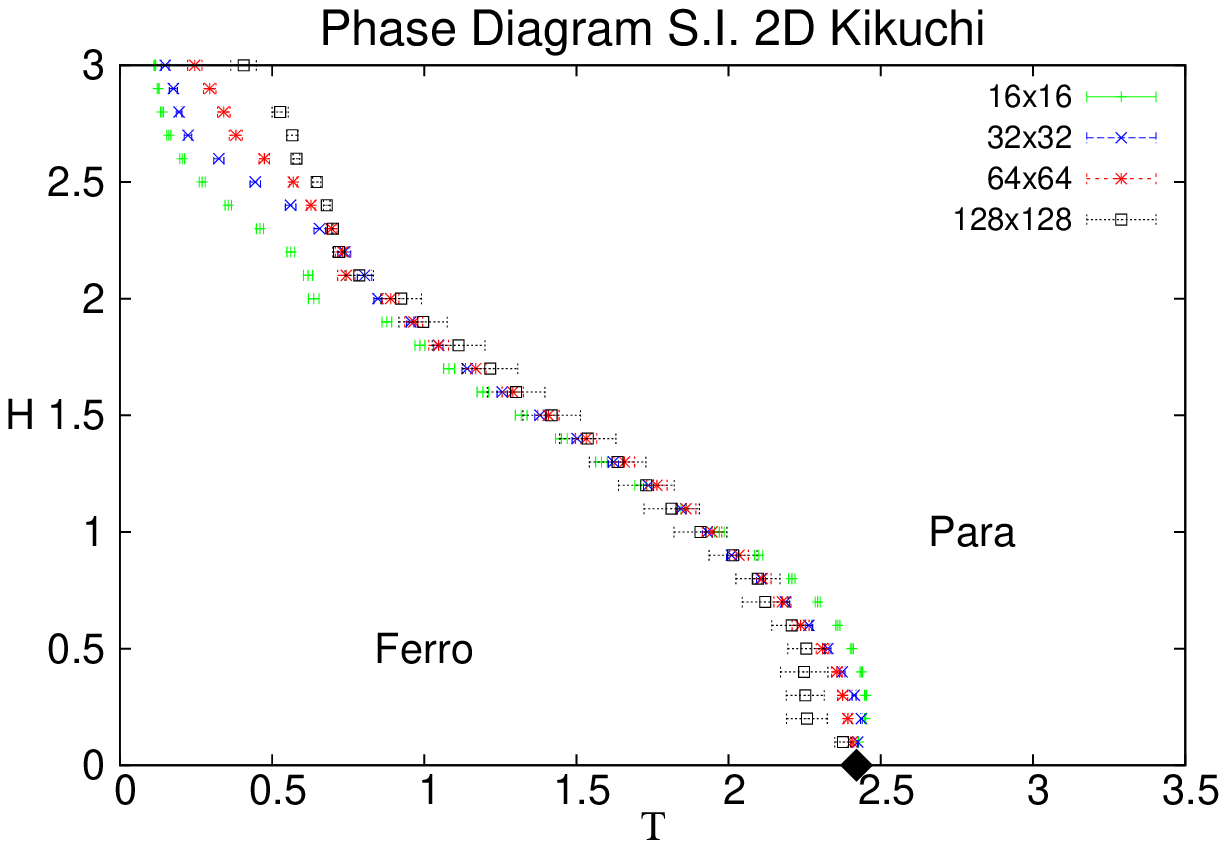}}
\subfloat[ Extrapolated $L\to \infty$ critical line. ]
{\label{fig:kikuchi_si_extrapolated}
 \includegraphics[keepaspectratio=true,angle=0,width=0.49\textwidth]{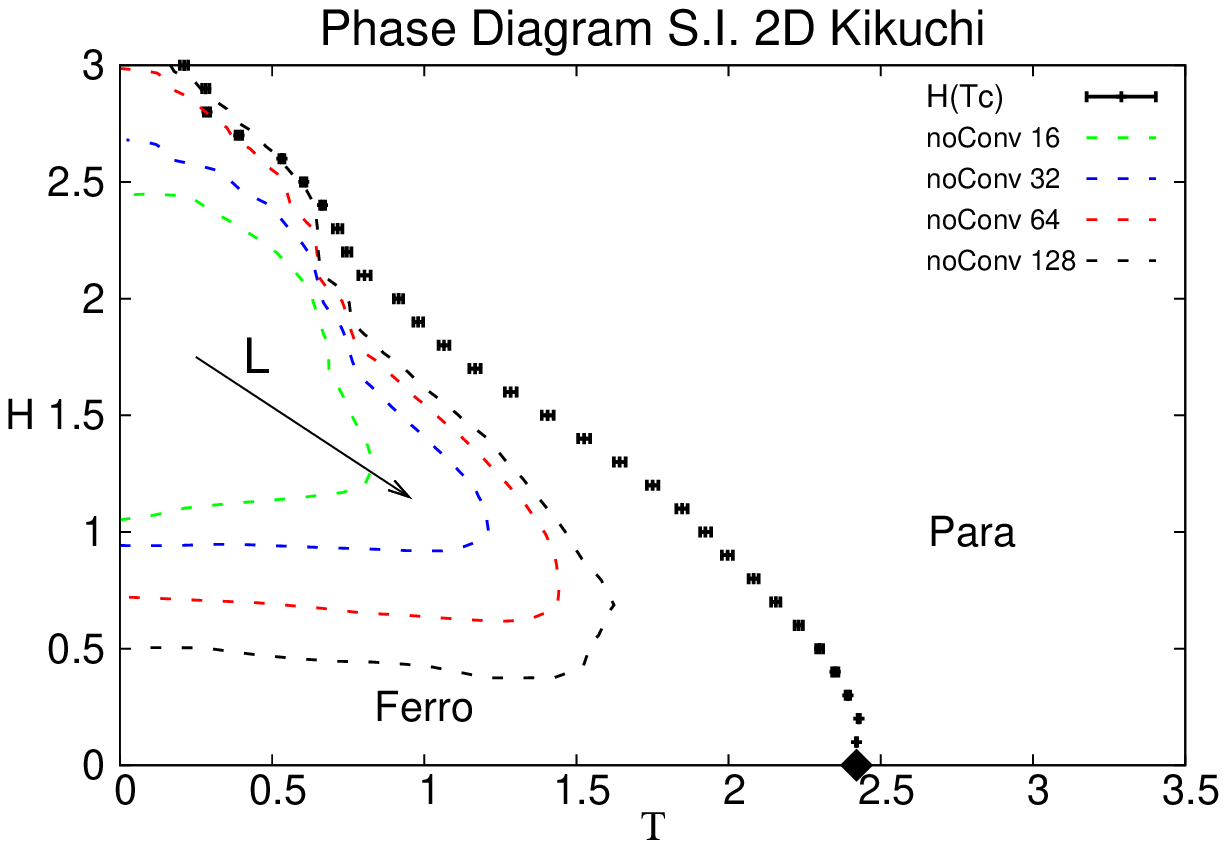}}
 \caption{Kikuchi(CVM) approximation. In part (b), for $\left(T,H\right)$ pairs inside the region
limited by dashed lines the algorithm does not converge. The non convergence region
seems to spread over the whole ferromagnetic phase as $L$ increases.}
\label{fig:kikuchi_si}
\end{figure}

Summarizing, in figures \ref{fig:bethe_si} and \ref{fig:kikuchi_si} we show the average phase diagram over many samples of RFIM with different
system sizes and the critical line after extrapolation. Our  Bethe and CVM calculations
predict the appearance of long range correlation in the magnetization of spins at low temperatures
and low external fields. This phase (thermodynamically incorrect for $L\to \infty$) will be called ferromagnetic.
In both cases the para-ferro border starts at the corresponding critical point of the ferromagnetic (non disordered, $H=0$) model.
This transition is at odds with the rigorous result proving that in $d\leq2$ even the smallest (random) external field
should destroy all chances of a long range order \cite{Imry_Ma_1,Imry_Ma_2,Bricmont}.

It is not a surprise that mean field approximations stabilize non relevant phases, as it
does in the Edwards-Anderson model \cite{GBPGF,average_mulet}. This is a general price we pay for
implicit factorization at certain levels of the correlations when writing the free energy approximated
by regions.

In the CVM plaquette approximation, we faced severe convergence problems that we solved partially by fixing the gauge
invariance of message passing equations\cite{GBPGF} as explained  in appendix \ref{ap:gauge}. In
the following, CVM approximation refers to this gauge fixed implementation of the message passing.
However, even in this case, there is an island of non convergence for low temperatures and intermediate values
of field (see \ref{fig:kikuchi_si}).

\begin{figure}
\subfloat[For a fixed $T$, $H(L)=C_H L^{-k_H}$ fits well the field value at the lower border of instable regions.]{\label{fig:kikuchi_non_convergence_henL}
\includegraphics[keepaspectratio=true,width=0.485\textwidth]{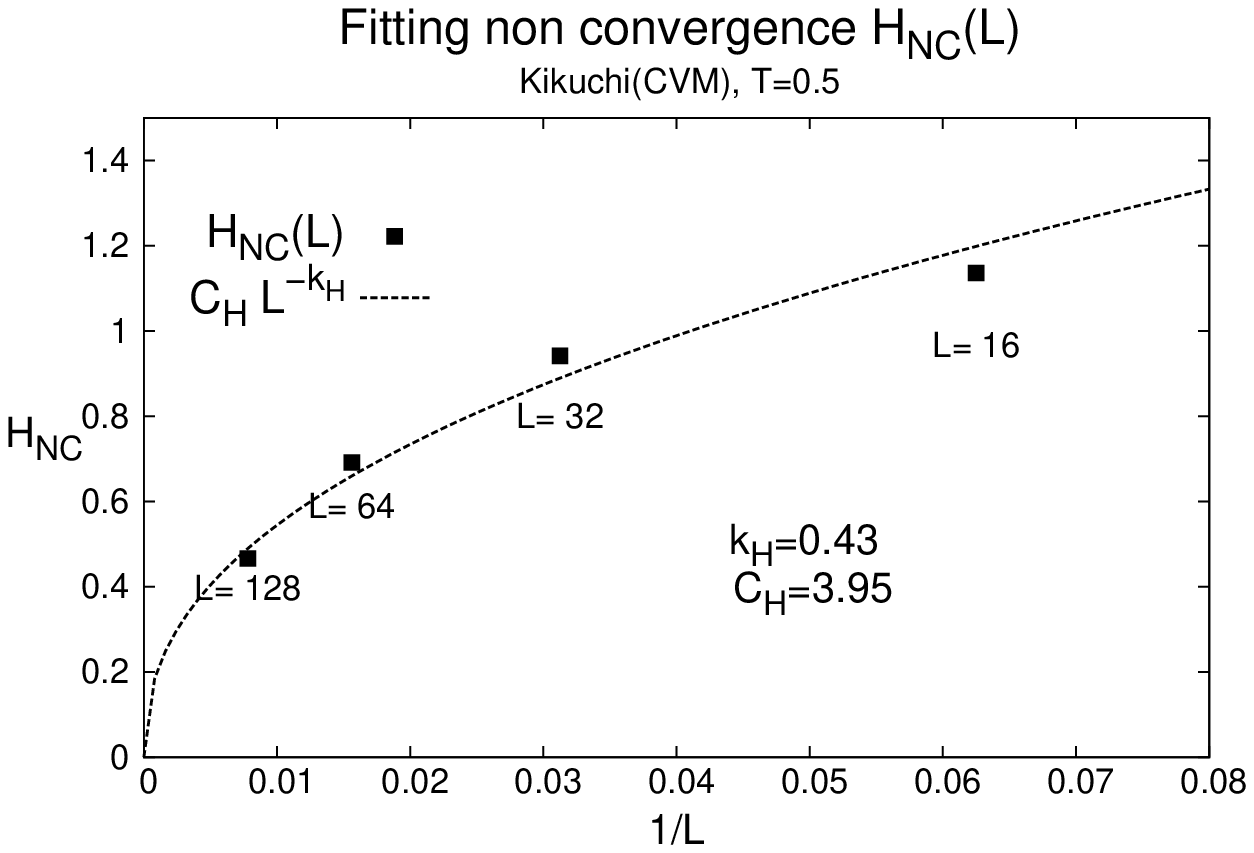}
}
\subfloat[For a fixed $H$, $T(L)=T_c- C_T L^{-k_T}$ fits well the temperature value at the right border of instable regions.]{\label{fig:kikuchi_non_convergence_TenL}
\includegraphics[keepaspectratio=true,width=0.485\textwidth]{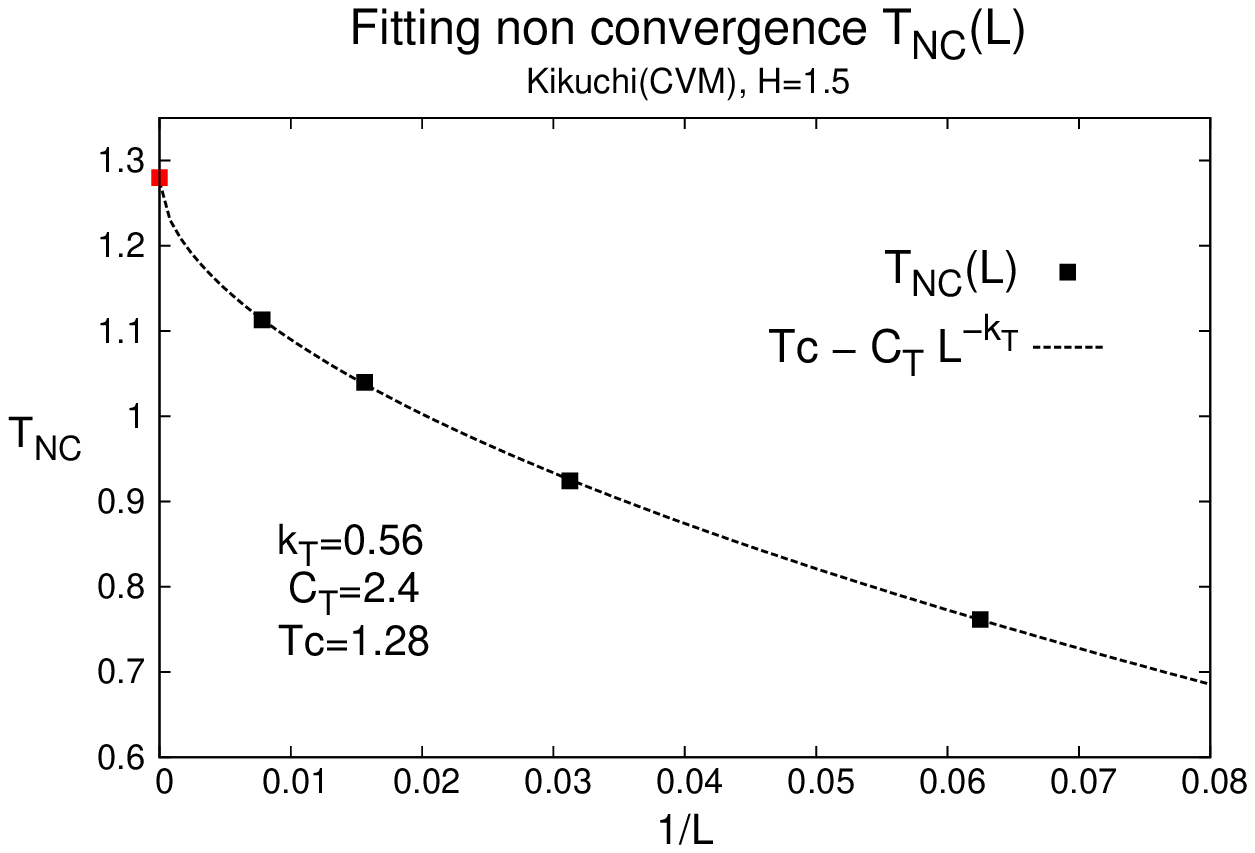}
}
 \label{fig:kikuchi_non_convergence}
\caption{Large $L$ behavior of the convergence frontier (see Fig. \ref{fig:kikuchi_si_extrapolated}). For $L \to \infty$,
the approximately flat lower border of the instable region approaches
the line $H=0$. On the other hand, the right border seems to approach the critical line.}
\end{figure}

This region of non convergence  grows with the system size.
In Fig. \ref{fig:kikuchi_si_extrapolated} we show, for different values of $L$ the border of the area where the probability of convergence drops abruptly.
As $L$ increases, the non convergence island apparently spreads all over the ordered phase.
To be more quantitative on this point we plotted for a fixed $T=0.5$ in Fig. \ref{fig:kikuchi_non_convergence_henL}
the value of $H=H_{\mbox{\tiny NC}}(L)$ at the frontier for each $L$. The functional dependence of this curve is very well
fitted  with the law $H(L)=C L^{-k}$. This suggests that indeed, for large enough lattices the non convergence region moves down to $H=0$.

Also, for a fixed $H=1.5$, we fitted in Fig. \ref{fig:kikuchi_non_convergence_TenL}
the temperature at the convergence border. The form of $T_{\mbox{\tiny NC}}(L)$ in this case agrees with $T_c-C L^{-k}$.
The limit $\displaystyle \lim_{L\to \infty} T_{\mbox{\tiny NC}}(L)=T_c=1.28$ is close to the average critical temperature found by GBP for the same
field intensity, $T_c(H=1.5)=1.40$. These results enable us to speculate that in the thermodynamic limit the difficulties in convergence would cover all the unphysical ordered phase. Moreover, this also suggests that at least for this model the convergence of GBP is linked to the para-ferro phase transition and may be useful in defining its location.

\subsubsection*{Multiple solutions: a one case study}

For many years there was a strong debate about whether a spin glass phase was present in an intermediate area of the
phase diagram just around the border of the para-ferro transition \cite{Imry_Ma_1,Imry_Ma_2,Bricmont}.
Although we now know that long correlation order is impossible in this case, such belief was supported by
perturbative and non perturbative replica field theory results\cite{DeDominicis,Brezin,Parisi_SG_sol_Replica}. On the other hand, although numerical Monte Carlo
simulations suffer from a strong slowing down near the transition, no evidence of a SG phase is found in either the
bimodal or the Gaussian version of the RFIM\cite{Newman_1,Middleton,Parisi_Fede_Montecarlo}.

\begin{figure}
\includegraphics[keepaspectratio=true,width=0.49\textwidth]{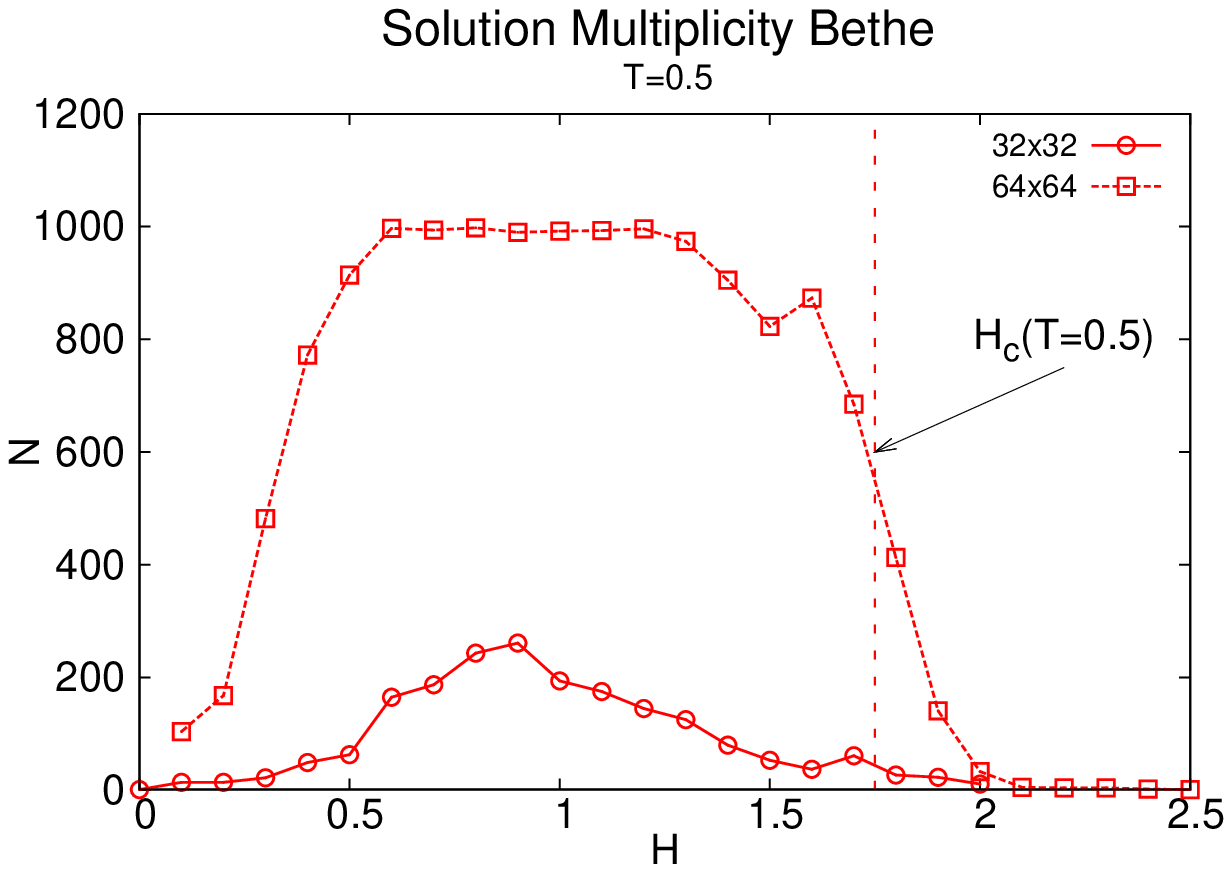}
\includegraphics[keepaspectratio=true,width=0.49\textwidth]{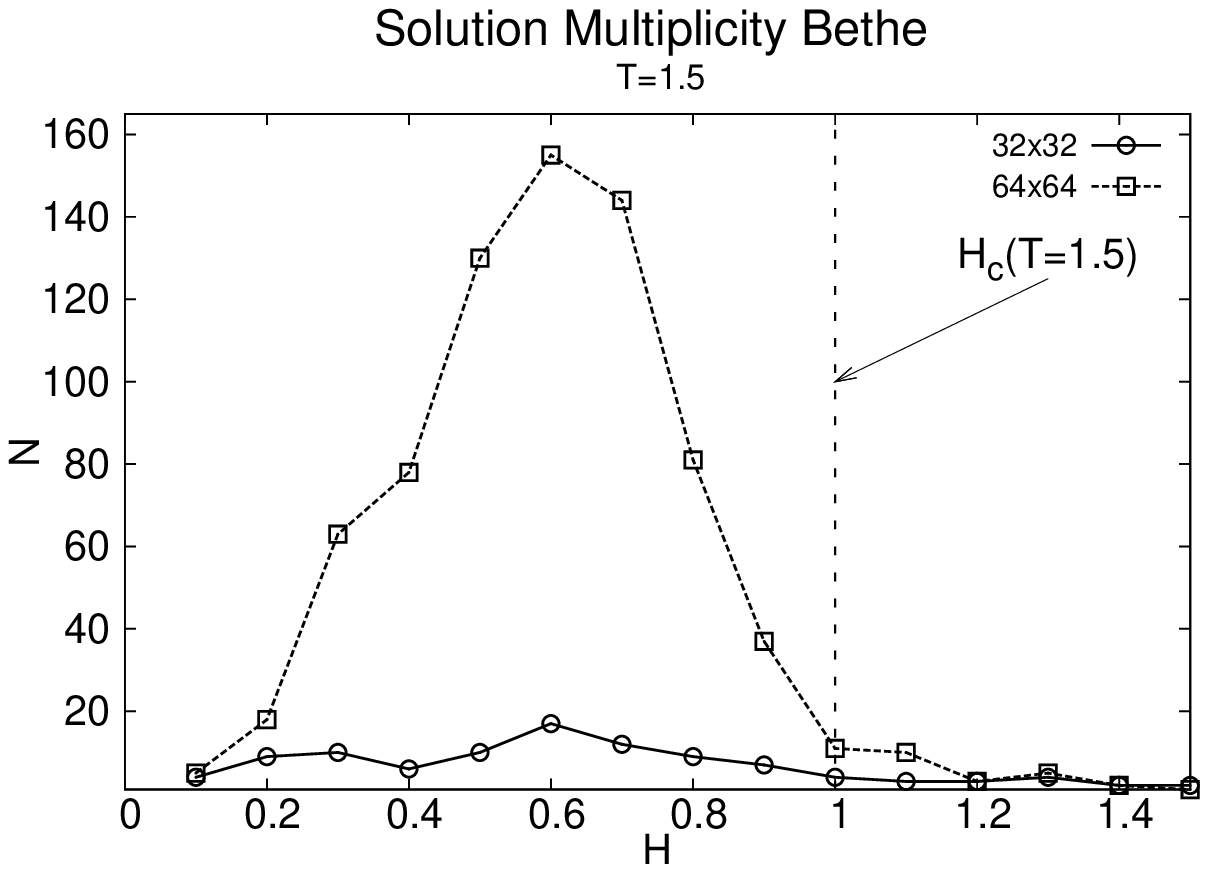}
\caption{Number of solutions found for two samples of different size as a function of field intensity. Multiple solutions appear for $H\lesssim H_c$,
within the ordered phase. For small $H$ the number of solutions reduces as for $H=0$ the solution is unique. The number
of stable fixed point for the BP equations increases with the size of the system and decreases with temperature.}
\label{fig:bethe_sols}
\end{figure}

It is consistent with such expectation the fact that Bethe approximation finds many different solutions
in the vicinity of the para-ferro transition. They are discerned easily by their different magnetizations.
In Fig. \ref{fig:bethe_sols} we show the number of different
solutions found in a system of $N=32\times 32$ and $N=64\times 64$  at $T=0.5$ and $T=1.5$. The message passing is
started from random initial conditions $10^3$ times for each value of the external field intensity. Changing the
external field means varying only its intensity while keeping directions on each site.
As expected. the larger the system the more solutions are found by the algorithm. Also, for lower temperatures
the number of solutions increase, as new local energy minima get stabilized.
Notice that multiplicity of solutions appears for $H \approx H_c(T)$ and below, within the long range order phase.
As a consequence, a proper definition of the critical field/temperature with a threshold in the global
magnetization is even more difficult than discussed previously.

\begin{figure}
      \includegraphics[keepaspectratio=true,angle=0,width=0.3\textwidth]{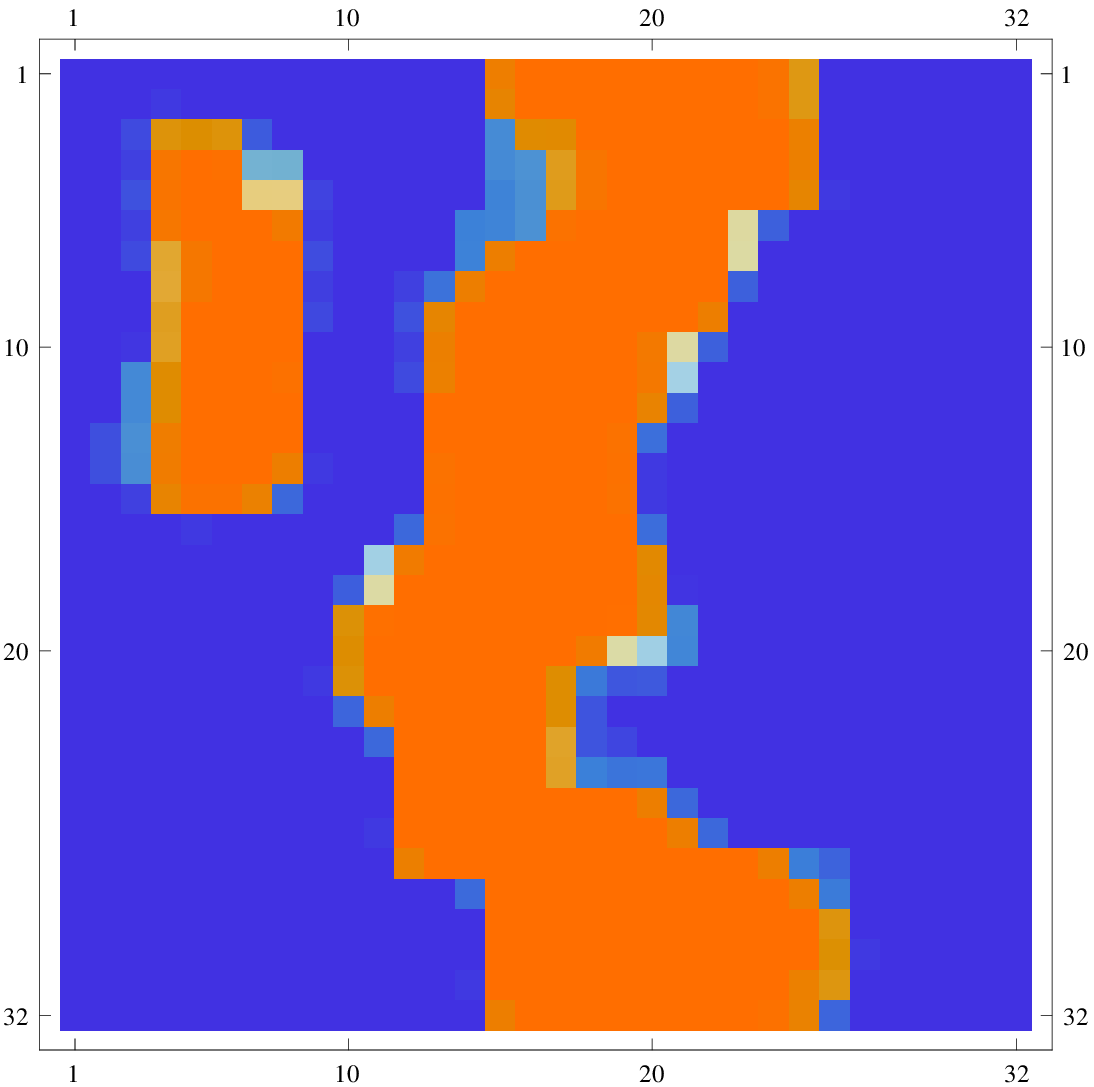}
      \includegraphics[keepaspectratio=true,angle=0,width=0.3\textwidth]{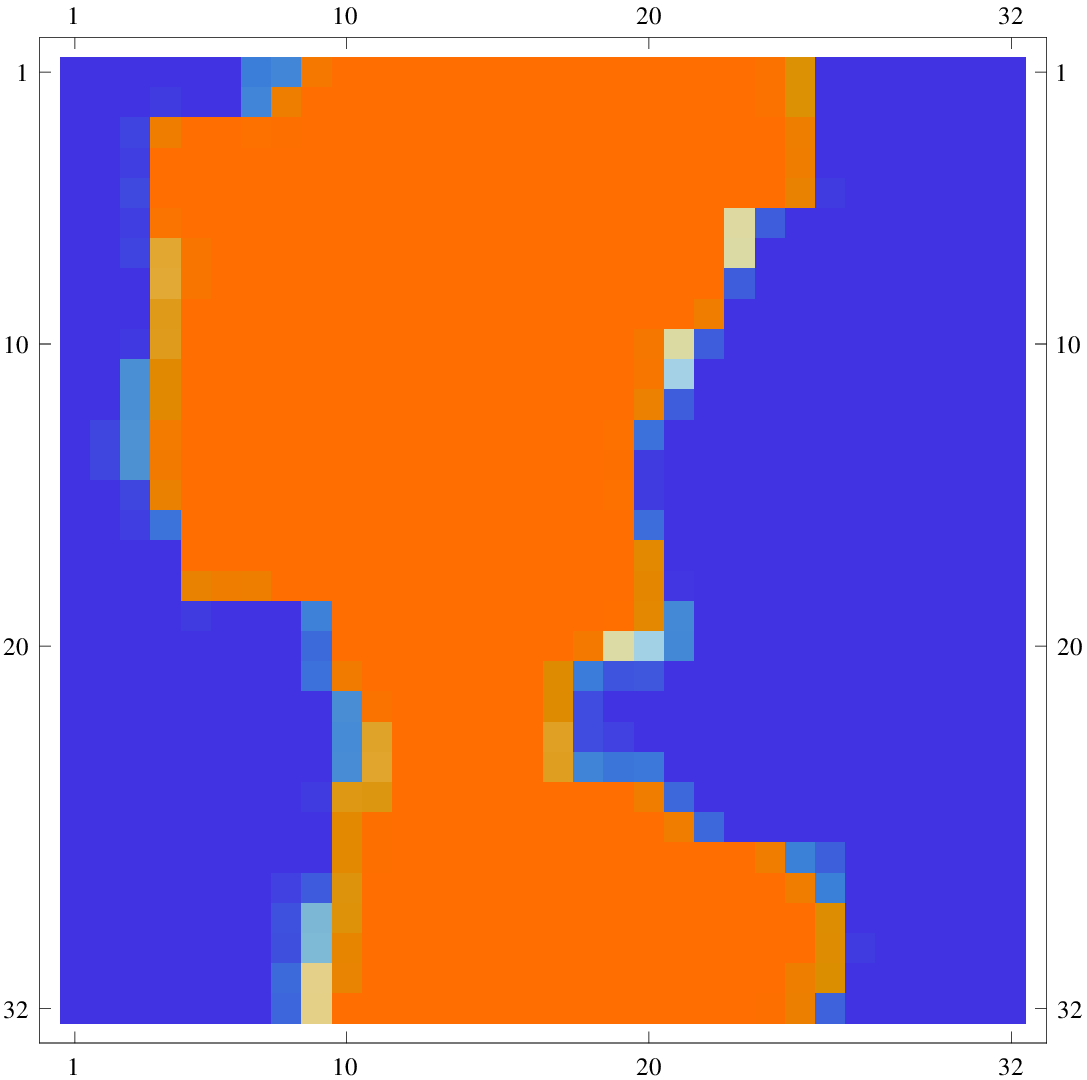}\\
      \includegraphics[keepaspectratio=true,angle=0,width=0.3\textwidth]{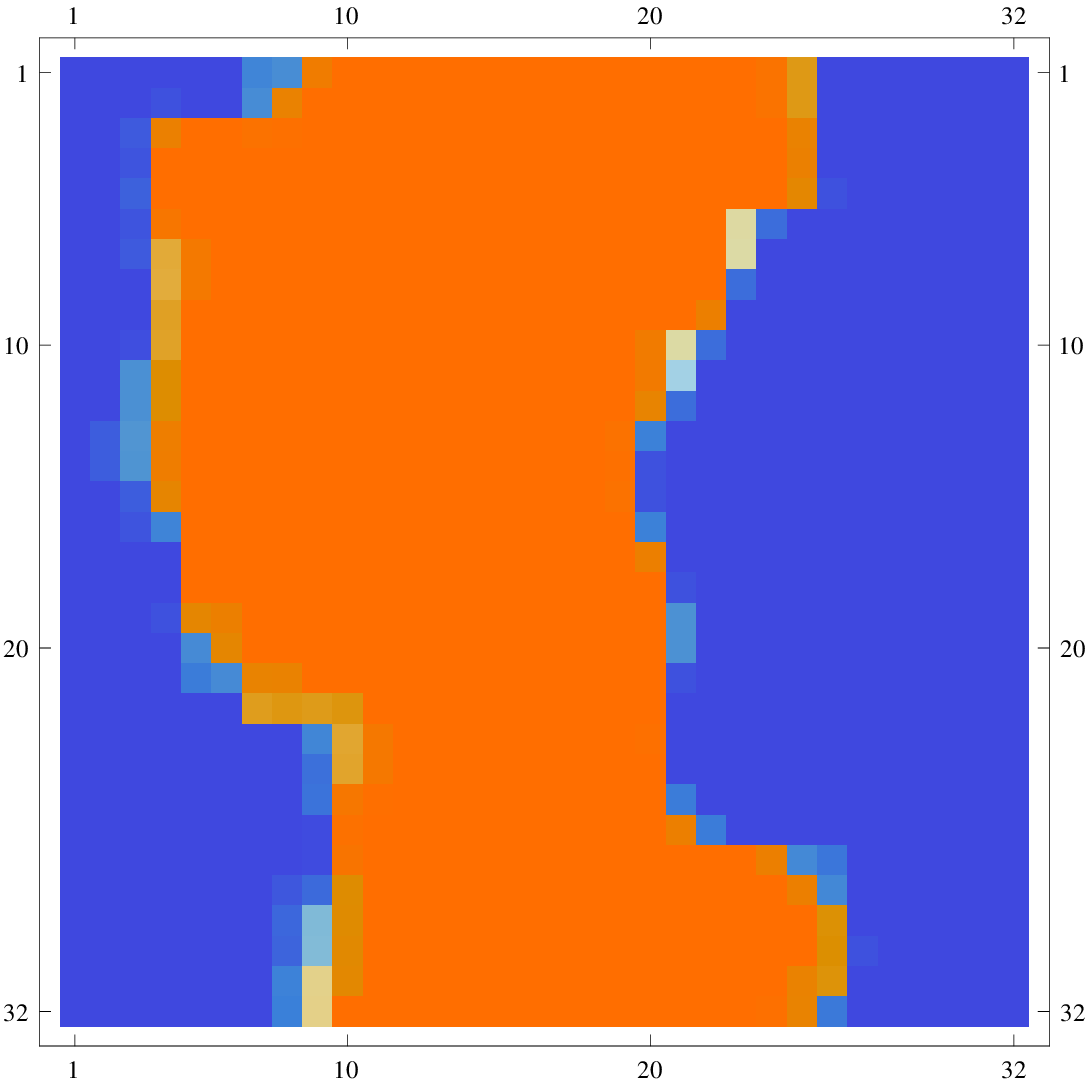}
      \includegraphics[keepaspectratio=true,angle=0,width=0.3\textwidth]{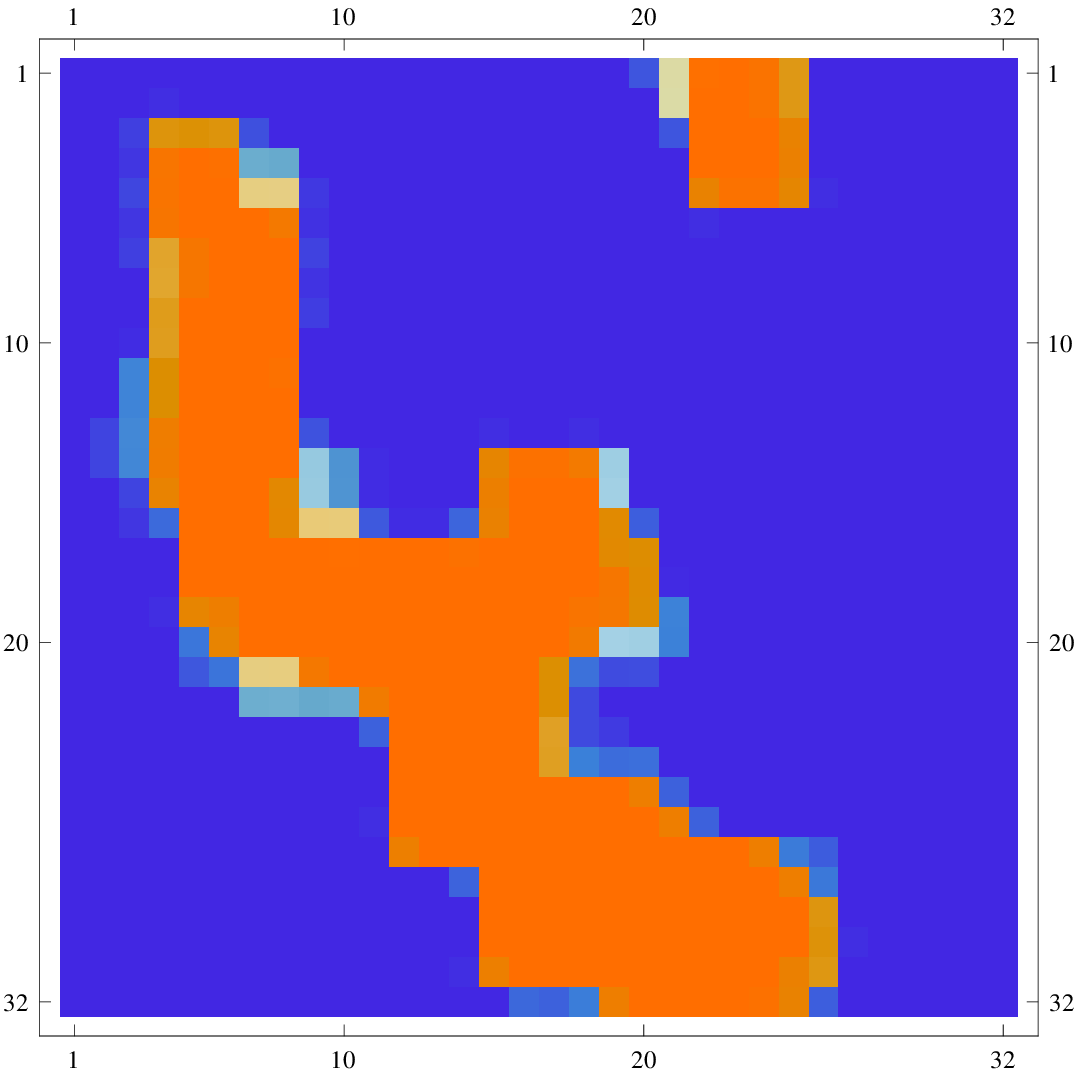}
\caption{These are four snapshots of the magnetization of a 32x32 system. Starting from different initial conditions for the same field values,
the BP algorithm finds different solutions characterized by domain formation.}
\label{fig:block_solutions}
\end{figure}

Different Bethe solutions are not completely uncorrelated. They differ in blocks of spins that switch
directions from one solution to the other, as shown in Fig. \ref{fig:block_solutions}. It has been
shown for other models, like the Edwards-Anderson in 2D, that these solutions might be connected with
the regions of the phase space in which the Monte
Carlo dynamics spend more time, or say in other words where the Boltzmann measure concentrates\cite{our_metastables}.
It is to be checked in future works whether this property holds for the RFIM too.

Furthermore, we wonder if this multiplicity of states is truly a sign
of a spin glass like behavior for finite lattice sizes, or also an artifact of the Bethe approximation. More precisely, we want
to check if it is also a feature of plaquette-CVM approximation. Surprisingly it is not. In most systems
studied, only one GBP fixed point was found. In some of them, two or at most three solutions were found.
When more than one solution is found, they are still related by
the flipping of entire blocks of magnetizations. The drastic reduction of fixed points for
plaquette-CVM is conspicuous, and might be interpreted as a sign of the
non thermodynamic relevancy of the many Bethe solutions. This is in accordance with the general belief that CVM
yields more accurate results than Bethe approximation.

\begin{figure}[hb!]
\includegraphics[keepaspectratio=true,width=0.49\textwidth]{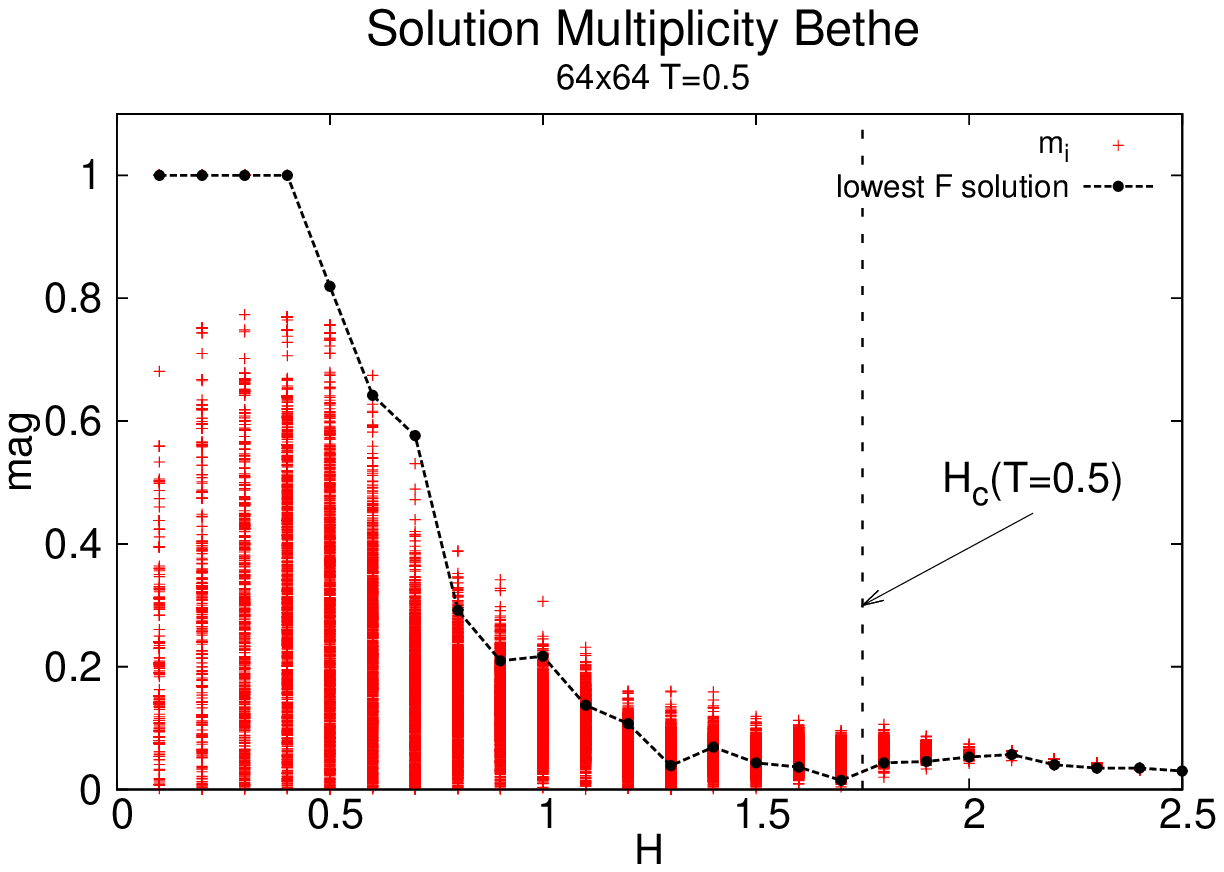}
\includegraphics[keepaspectratio=true,width=0.49\textwidth]{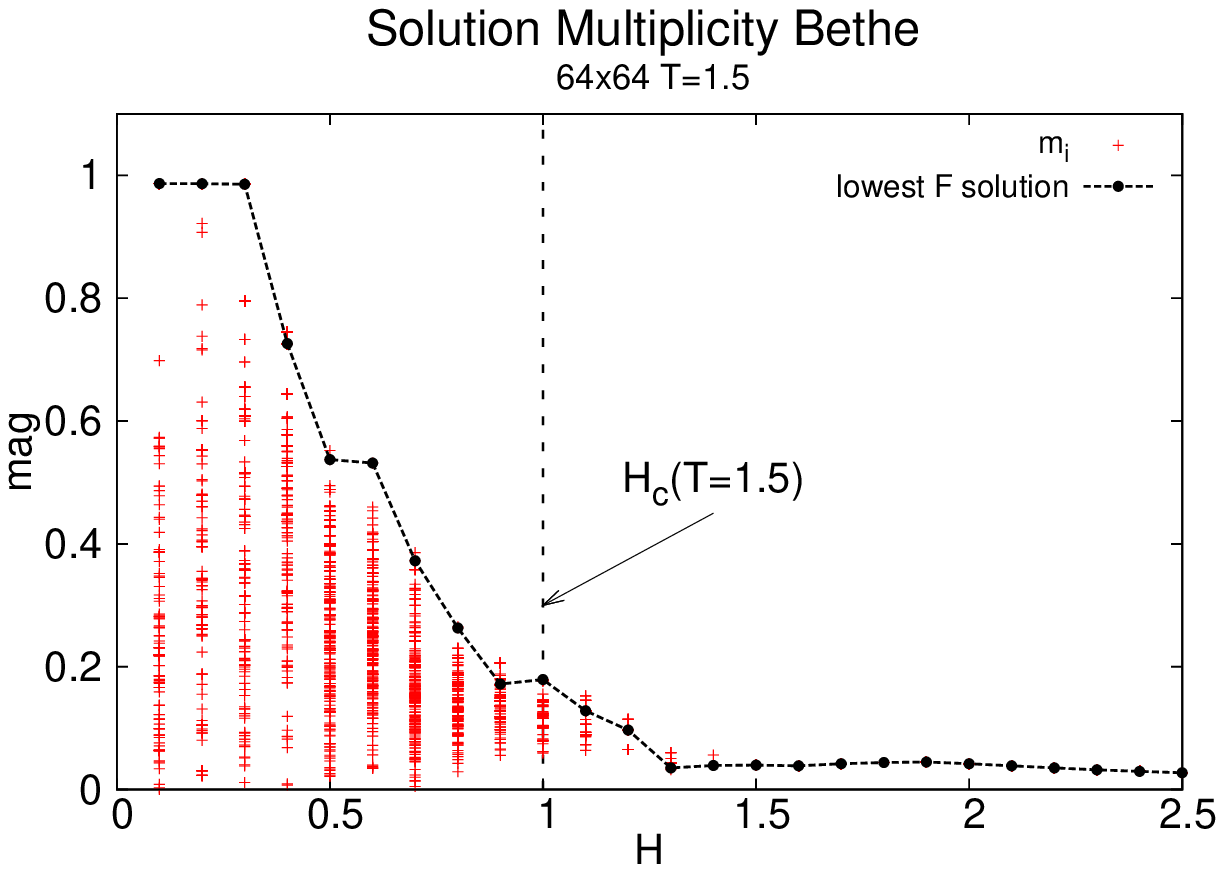}
\caption{For a given 64x64 system, different magnetizations are obtained for each $H$ value when running BP with various initial
conditions. The solution with the lowest free energy is drawn with a solid line. This line should be useful in defining the transition more
accurately. The $H_c$ values correspond to the temperature of each plot and are taken from the results shown in Fig. \ref{fig:bethe_si}.}
\label{fig:bethe_sols_mag_free}
\end{figure}

This detailed study of the Bethe and CVM approximations shows the difficulties affronted in defining a critical
line in the random field Ising model. At variance with the Ising ferromagnet, where the Bethe and CVM
approximations give a clear zero magnetization in the paramagnetic phase\cite{OurRCF}, here there are non zero
magnetizations at every temperature and field, and the onset of long correlation is further obscured
by the appearance of  many solutions. It would be more justified if one could define the critical line
studying a threshold magnetization value for the solution with lowest free energy at every $(T,H)$.
The different magnetization values for a 64x64 sample are plotted in Fig. \ref{fig:bethe_sols_mag_free}.
Solutions with the lowest free energy are joined with a solid line. Applying the threshold criteria to this
line should improve the quality of the results. However, this requires running message passing thousands of times from
random initial conditions for each system, and then the average over thousand of systems. This is a highly resource-consuming task.
Instead, we turned to the prediction of a critical line by average case calculations of both Bethe and plaquette-CVM approximations.

\section{Average case predictions}
\label{sec:Ave}

There are two different procedures to analyze the average case solution
of CVM approximations, one more formal relying on the replica method \cite{tommaso_CVM}
and quite involved, and another more intuitive based on population dynamics \cite{average_mulet}.
In the case of the Bethe approximation both procedures are equivalent, the latter
being a numerical solution of the integral equation appearing in the former \cite{MP1}.
A detailed explanation of both procedures can be found in \cite{average_mulet}, and we will
only describe them shortly.

Replica-CVM is a formal way of averaging the disorder in the partition function of
a region graph approximated disordered model, based on the replica trick \cite{tommaso_CVM}.
As usual in replica method, the average
over the disorder couples the replicas, and some ansatz is needed to take the $n\to 0$ limit.
Generalizing the approach in \cite{Monasson_param,kabashima} to the plaquette-CVM case,
this ansatz is a parametrization of messages $m$ and $M$ in terms of two types of
field distributions $q(u)$ and $Q(U,u_1,u_2)$ respectively:
\begin{eqnarray}
\label{eqn:parametric_m}
 m(\sigma_i)&\propto& \int du q(u) \exp \left[\beta u \sum_{a=1}^n \sigma_i^a
 \right]
\\
\label{eqn:parametric_M}
M(\sigma_i,\sigma_j)&\propto& \int dU du_i du_j Q(U,u_i,u_j)\exp  \left[\beta
U \sum_{a=1}^n
\sigma_i^a\sigma_j^a
+ \beta u_i \sum_{a=1}^n \sigma_i^a + \beta u_j \sum_{a=1}^n \sigma_j^a
  \right]
\end{eqnarray}
Minimizing the replicated free energy in terms of these distributions, we
obtain two self consistent equations for $q(u)$ and $Q(U,u_1,u_2)$:
\begin{eqnarray}
\nonumber
q(u)&=&\int  \prod_i^k dq_i \prod_\alpha^p dQ_\alpha \langle \delta(u-\hat
u(\#))
\rangle_h\\
\label{eqn:update_Qq}
R(U,u_a,u_b)&\equiv&\int du_i du_j Q(U,u_i,u_j) q(u_a-u_i) q(u_b-u_j)=
\\
\nonumber
&=& \int  \prod_i^K dq_i \prod_\alpha^P dQ_\alpha \langle \delta(U-\hat
U(\#))  \delta(u_a-\hat
u_a(\#)) \delta(u_b-\hat u_b(\#)) \rangle_h
\end{eqnarray}
with $\hat u(\#)$ and $ \hat U(\#)$ functions defined in (\ref{eqn:hat_u}) and
(\ref{eqn:hat_Uuu}) and $$dq_i \equiv q(u_i) du_i \mbox{\hspace{1cm}} dQ_\alpha
\equiv Q_\alpha(U,u_i,u_j) dU du_i du_j$$
All the thermodynamics is encoded into the functions $q$ and $Q$. The solution of
this system is technically involved. Standard population dynamics is not easily
applicable because of the need of numerical deconvolution methods (like Fourier transform)
to extract $Q$ from the convolution in the left side of Eq(\ref{eqn:update_Qq}). In the case
of zero external field and high $T$, however,  there is the trivial solution $q(u)= \delta(u)$,
and $Q(U,u_1,u_2) = Q(U)\delta(u_1)\delta(u_2)$, with $Q(U)$ satisfying a simplified version
of the integral equation (\ref{eqn:update_Qq}) (see \cite{average_mulet} for details).

We reproduce the stability analysis of the paramagnetic solution for the RFIM, knowing in advance, that $q(u) = \delta (u)$ is
not a good guess in the $H \gg 0$ zone, since at any finite temperature
the local magnetizations are non zero, and therefore, messages should be non trivial even
in the paramagnetic phase. Around the paramagnetic solution the self consistent
integral equations (\ref{eqn:update_Qq}) can be linearized for the first moments
of the distributions $q(u)$ and $Q(U,u_1,u_2)$, and the critical temperature is
defined as the moment when the system becomes singular \cite{average_mulet}.

In the simplest case of the Bethe approximation, only an equation like the first one in
(\ref{eqn:update_Qq}) appears (excluding the presence of the $Q(u,u_1,u_2)$ functions),
and it can be solved again in terms of the moments around $q(u)=\delta(u)$, or numerically
as a fixed point dynamic over a population of fields $u$ representing $q(u)$ \cite{MP1}.

\begin{figure}[!htb]
 \includegraphics[keepaspectratio=true,width=0.45\textwidth]{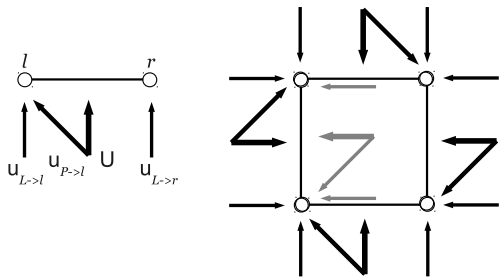}
\caption{Messages in the population dynamics scheme. In order to preserve correlations, plaquette to link
messages are updated and stored together with the parallel link to spin messages as shown in the left part of
this figure. Each iteration step samples the surroundings of a plaquette and generates interior messages (see right side).}
\label{fig:pop_CVM}
\end{figure}

This population sampling method can be extended to the CVM approximation in a less trivial way. A
representation of $Q(U,u_1,u_2)$ with a population of fields $(U,u_1,u_2)$ is not enough to attempt
a numerical solution of equation  (\ref{eqn:update_Qq}), since it is not trivial how to
deconvolve $Q$ in the left hand side of (\ref{eqn:update_Qq}). Still, you can pretend that a
random sampling of $u$ and $(U,u_1,u_2)$ populations and the evaluation of message passing
equations (\ref{eqn:hat_u}) and (\ref{eqn:hat_Uuu}) with random local external fields $\tilde h_i$
represents a randomized version of the actual message passing occurring in the system, and
therefore represents an {\it average} system.

In the single instance case, for every update step of $(U,u_1,u_2)$ messages, it is necessary to update simultaneously the two ($u_L$)
messages that enters in the left hand side of (\ref{eqn:updateplaqtolink})\cite{YFW05}. It is prudent then to respect
this correlation in the random sampling procedure. In order to do so, a random object containing a $(U,u_1,u_2)$ and two $u_L$
is defined (see Fig. \ref{fig:pop_CVM}, left). It is possible to generate a new set of $(U,u_1,u_2)$ and $u_L$ by sampling four
of these objects from a large population and using them as the messages acting
on a plaquette (see Fig. \ref{fig:pop_CVM}, right). This newly generated random object is
added back to the population at the end of each iteration step.

\begin{figure}
 \includegraphics[keepaspectratio=true,width=0.7\textwidth]{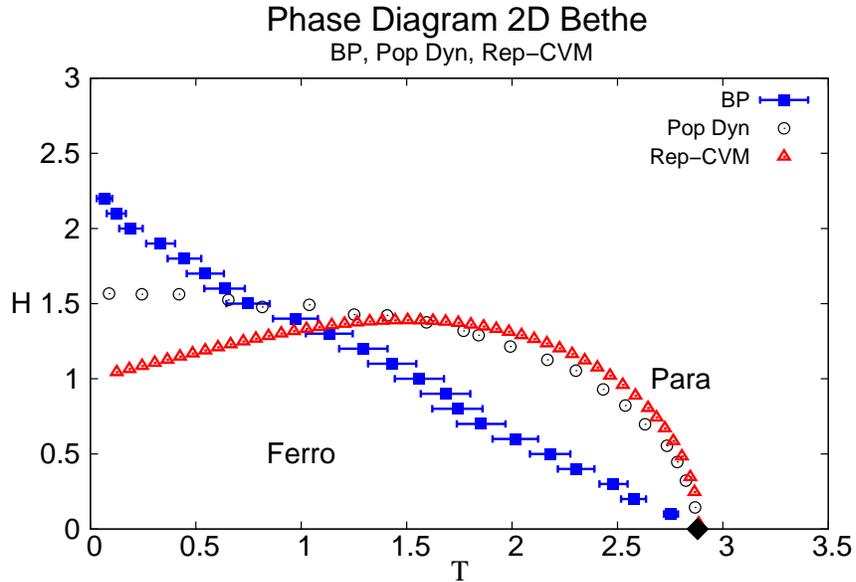}
\caption{Comparing results for the Bethe approximation. In this case, the single instance results differ from the two average
case prediction. The fact that for a given $H$, the critical temperature is generally lower than the average calculation
is a consequence of the domain formation phenomena}
\label{fig:bethe_crit_lines}
\end{figure}

In the population dynamic method we will define the critical temperature as the moment in which
the population of fields develop a global magnetization. This means that the balance between
positive and negative $u$ fields is skewed to one side, signaling the appearance of long range
order.

\begin{figure}
 \includegraphics[keepaspectratio=true,width=0.7\textwidth]{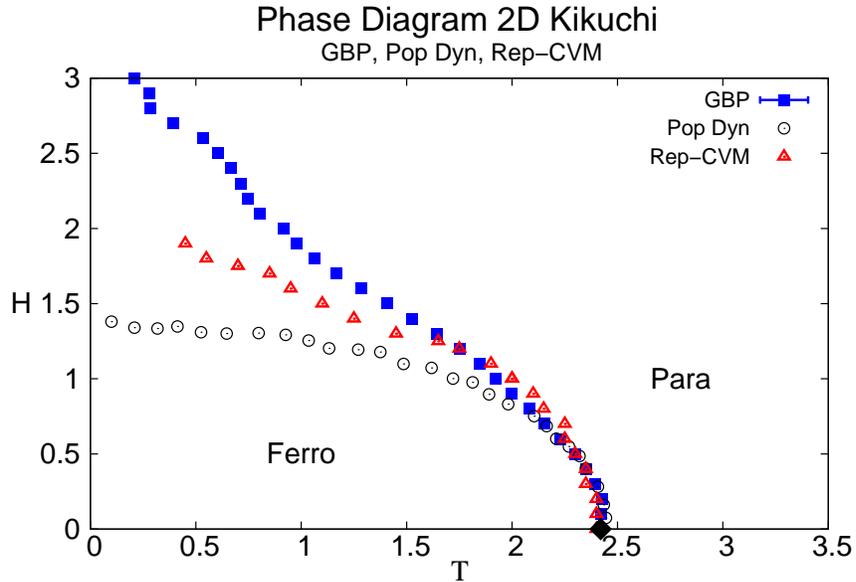}
\caption{The GBP approximation match quite well for small $H$ values with average case predictions. Calculations for the
Replica CVM are valid only for weak magnetic field, whereas, for high intensities, approximations made in calculations break down. Error bars
for the GBP graphic fall within the symbol shape.}
\label{fig:cvm_crit_lines}
\end{figure}

In Fig. \ref{fig:bethe_crit_lines} we show the critical lines obtained by examine
single instances of RFIM in 2D with the Bethe approximation (same data as in Fig. \ref{fig:bethe_si_extrapolated}),
and the results obtained in average case by both procedures: stability analysis of linearized equation and population dynamics.
Remember that for the Bethe case these two methods attempt to solve the same integral equation. Therefore, had the
stability analysis been performed to all orders, it would have matched population dynamics results. Nevertheless, for small
$H$, a good agreement is observed in Fig. \ref{fig:bethe_crit_lines}. In the limiting case $H=0$ all results  agree on
the same critical temperature, which can be found accurately\cite{OurRCF}. In presence of disorder, on the other hand, the single instance
critical line predicts for the same field intensity a transition temperature which is lower than the average case values.
This divergence is closely related to the domain formation picture, because the system needs lower temperatures to be completely
magnetized.

The same comparison is done in Fig. \ref{fig:cvm_crit_lines}  for the plaquette-CVM approximation. Again, at low $H$ both
average case calculations coincide. Furthermore, they coincide also with the results for the phase diagram obtained within
the single instance scenario.

At this point it is interesting to compare these results for CVM with previous predictions for the EA model \cite{average_mulet}.
Unfortunately the replica stability analysis in RFIM can only be accurate in the low field zone, restricting our analysis to this part
of the diagram. While in the 2D EA model the population dynamic and the replica stability give different transitions temperatures,
the first one related to the appearance  of non paramagnetic GBP solutions and the second pointing the lack of convergence of the
message passing, in RFIM both methods seems to coincide (in the low H region). They also coincide with the appearance of magnetized
solutions in single instances, and asymptotically with the lack of convergence for $L\to\infty$. In this sense the relation of single
instance behavior of GBP and both average case methods is connected in a similar fashion as in 2D EA. The rest of the diagram
(higher $H$) is harder to analyze, since replica stability is a priori wrong, and population dynamics gets into the non convergent
region of single instances.

\section{Conclusions}
\label{sec:Conc}

The Random field Ising model is a paradigmatic disordered model in statistical mechanics whose full comprehension in
general dimensions resists analytic methods and poses major simulation difficulties. We have studied two approximations
to the free energy of this model in 2D, namely Bethe and plaquette-CVM approximations, both in single instances
and in the average case.

The qualitative panorama of both approximations is consistent. BP and GBP on finite size instances and average case calculations predict
 transitions to an ordered phase in a region of low temperatures and low fields. Although it is known that this phase is thermodynamically
 unstable, the prediction of long range order in 2D is not a surprise, since  mean field like approximations tend to stabilize
ordered phases. However, Bethe and CVM differ in many aspects. The Bethe approximation predicts a critical line that is above the one of
 the plaquette-CVM, which is consistent with the expectation that a more precise approximation yields results closer to reality
(no critical line at all). On single instances the Bethe approximation converges to many solutions and GBP to only one or at
most a few ones related by flipping large spin clusters. Moreover, our results suggest that GBP does not converge for
large enough lattices and  more important, at least for small values of the field $H\sim 0$, the non-convergence  seems to coincide
with the para-ferro transition temperature predicted by the population dynamic and the replica stability analysis. As in the 2D EA model,
the behavior of GBP is closely connected with the average case prediction.

\section{Appendix: Gauge invariance of GBP equations}
\label{ap:gauge}

GBP equations enforce the correct marginalization of the expectations
(beliefs) at each level onto the expectation at
its children levels, in a region graph approximation to the free energy of
a model. As discussed in \cite{yedidia},
each marginalization requires a Lagrange multiplier, and the self
consistency of this multipliers becomes the usual
message passing algorithm.

When implementing GBP in single instances, it makes no damage if some of
the constraints are forced more than once.
For instance, if a plaquette's belief $b(\sigma_1,\sigma_2,\sigma_3,\sigma_4)$ is forced to
marginalize onto two of its links $b(\sigma_1,\sigma_2)$
and $b(\sigma_4,\sigma_1)$, then the marginals of these two beliefs must be
consistent on variable $\sigma_1$.
Therefore, if one of them, lets say $b(\sigma_1,\sigma_2)$ is forced to marginalize
on the belief $b(\sigma_1)$, then it is unnecessary
to force the other to marginalize on it, since it inevitably does.

\begin{figure}
 \includegraphics[keepaspectratio=true,angle=0,width=0.5\textwidth]{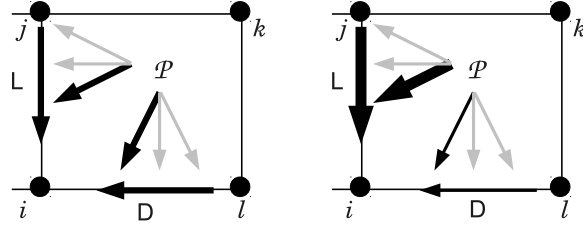}
\caption{One of the gauge invariances of CVM approximation. Since fields are under determined, some can be raised(thicker arrows) and
others diminished (thinner arrows) without altering the total field acting on a site. Eliminating this free modes by fixing some fields can
help improving convergence properties of the algorithm. The procedure of fixing the gauge does not affect the belief or other relevant magnitudes.}
\label{fig:invariance_u}
\end{figure}

This means that the general prescription for implementing GBP in
\cite{yedidia} is redundant, and the introduced messages
or their equivalent cavity fields are under-determined. In \cite{GBPGF} we
discussed this fact for the Edwards Anderson 2D
model, and showed a way to fix the gauge invariance in the cavity fields
that appears as a result of their indeterminacy.
In Fig. \ref{fig:invariance_u} we show schematically one of the gauge
invariant transformations of the messages. Two of
the messages pointing to spin on bottom-left can be raised by the same
arbitrary amount, and the two others reduced in the
same amount without altering the fixed point of the equations.

\begin{figure}
 \includegraphics[keepaspectratio=true,angle=0,width=0.5\textwidth]{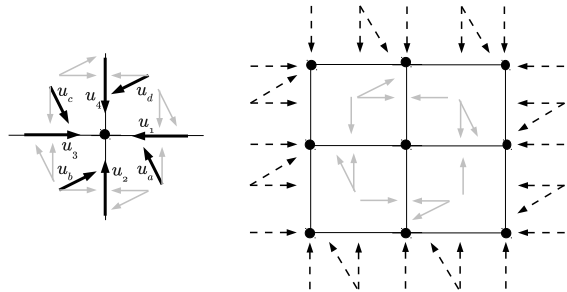}
\caption{Fields shown in bold face in this figure enter linearly into each other update equation. This fact
can be exploited to update them in a simultaneous and coherent way. In addition to the gauge fixing this heuristic
helped us improving convergence near the transition line.}
\label{fig:invariance_u_mat}
\end{figure}

Fixing the gauge does not alter the type or the number of solutions found
for the beliefs (which is the physically meaningful
quantity, not the fields or messages). However, it can help
gaining some more convergence in single instances, and
furthermore, becomes necessary to make a correct population dynamics
representation of the average case. We decided to fix the
gauge by setting to zero one of the fields acting over spins in the
plaquette-to-link message $(U,u_1,u_2)$, as shown in
Fig. \ref{fig:invariance_u_mat}. Furthermore, since all the bold faced
fields in this figure participate in each other
equations in a linear way, we solved the set of linear equations in each
message passing step, therefore moving all these
fields in a consistent way towards agreement\cite{GBPGF}.

\begin{figure}
 \includegraphics[keepaspectratio=true,angle=0,width=0.5\textwidth]{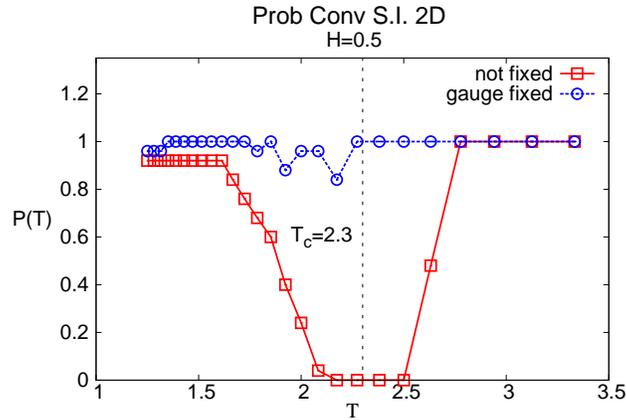}
\caption{Comparing GBP  with and without gauge fixing in a sample. Probability of convergence is plotted as a function of temperature
for a fixed external magnetic field. Although near the transition both  variants get instable, the one with the fixed gauge converges
much better.}
\label{fig:cvm_si}
\end{figure}

This mixed strategy of fixing the gauge and a consistent updating of all
linearly dependent fields showed to improve the convergence of the GBP in
the 2D Edwards Anderson model\cite{GBPGF}. In Fig. \ref{fig:cvm_si}
we show that this is also the case in the random field Ising model. Most
of the convergence problem appearing near the transition line
from para to long range ordered phase disappears if the GBP is implemented
with these two prescriptions. However, the convergence problems
appearing within the long range ordered phase, did not disappear, and a
growing island of non convergence still exists for low temperatures
and fields, as discussed in the article.

\newpage

\bibliography{bibliografia}
\end{document}